\documentclass[USenglish,oneside,twocolumn]{article}

\usepackage[utf8]{inputenc}
\usepackage[big]{dgruyter_NEW}
\usepackage{comment}
\usepackage{stmaryrd} 
\usepackage{mathtools} 
\usepackage{multirow}
\usepackage[dvipsnames]{xcolor}

\usepackage{pifont}
\usepackage{graphicx}
\usepackage{tikz}
\usepackage{siunitx}
\usepackage{fontawesome5}
\usepackage[all]{nowidow} 

\usepackage{tcolorbox}
\usepackage{hyperref}
\usepackage{xcolor}

\addto\extrasUSenglish{ 

}

\newcommand{\greencheck}{{\color{OliveGreen}\ding{51}}}
\newcommand{\redcheck}{{\color{red}\ding{55}}}

\newcommand{\cactus}{\texttt{CaCTUs}}

\newcommand{\reslatency}{\SI{2}{s}}
\newcommand{\resfps}{\SI{10}{fps}}
\newcommand{\resquality}{\SI{480}{p}}

\newcommand{\hypertargetblue}[2]{\textbf{\textcolor{blue}{\hypertarget{#1}{#2}}}}

\newcommand{\mr}[1]{[MR.#1]}
\newcommand{\re}[2][]{[R#2.#1]}

\newif{\iffinal}    
\finaltrue 
\newif{\ifreviews}
\reviewsfalse 
\newif{\ifrevision}
\revisionfalse 

\DOI{foobar}

\cclogo{\includegraphics{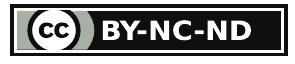}}
 
\begin{document}

\iffinal
  \author*[1]{Yohan Beugin}
  \author[2]{Quinn Burke}
  \author[3]{Blaine Hoak}
  \author[4]{Ryan Sheatsley}
  \author[5]{Eric Pauley}
  \author[6]{Gang Tan}
  \author[7]{Syed Rafiul Hussain}
  \author[8]{Patrick McDaniel}

  \affil[1]{The Pennsylvania State University, E-mail: yohan@beugin.org}
  \affil[2]{The Pennsylvania State University, E-mail: qkb5007@psu.edu}
  \affil[3]{The Pennsylvania State University, E-mail: bhoak@psu.edu}
  \affil[4]{The Pennsylvania State University, E-mail: sheatsley@psu.edu}
  \affil[5]{The Pennsylvania State University, E-mail: epauley@psu.edu}
  \affil[6]{The Pennsylvania State University, E-mail: gtan@psu.edu}
  \affil[7]{The Pennsylvania State University, E-mail: hussain1@psu.edu}
  \affil[8]{The Pennsylvania State University, E-mail: mcdaniel@cse.psu.edu}
  
\else
  \author*[1]{Corresponding Author}
  \author[2]{Second Author}
  \author[3]{Third Author}
  \author[4]{Fourth Author}

  \affil[1]{Affil, E-mail: email@email.edu}
  \affil[2]{Affil, E-mail: email@email.edu}
  \affil[3]{Affil, E-mail: email@email.edu}
  \affil[4]{Affil, E-mail: email@email.edu}
\fi
   \title{\huge Building a Privacy-Preserving Smart Camera System}

  \runningtitle{Building a Privacy-Preserving Smart Camera System}


\begin{abstract}
{
Millions of consumers depend on smart camera systems to remotely monitor their homes and businesses. However, the architecture and design of popular commercial systems require users to relinquish control of their data to untrusted third parties, such as service providers (e.g., the cloud). Third parties therefore can (and in some instances have) access the video footage without the users' knowledge or consent---violating the core tenet of user privacy. In this paper, we present \cactus{}, a privacy-preserving smart Camera system Controlled Totally by Users. \cactus{} \textit{returns control to the user}; the root of trust begins with the user and is maintained through a series of cryptographic protocols, designed to support popular features, such as sharing, deleting, and viewing videos live. We show that the system can support live streaming with a latency of \reslatency{} at a frame rate of \resfps{} and a resolution of \resquality{}. In so doing, we demonstrate that it is feasible to implement a performant smart-camera system that leverages the convenience of a cloud-based model while retaining the ability to control access to (private) data.

}
\end{abstract}

  \journalname{Proceedings on Privacy Enhancing Technologies}
  \DOI{Editor to enter DOI}
  \startpage{1}
  \received{..}
  \revised{..}
  \accepted{..}

  \journalyear{..}
  \journalvolume{..}
  \journalissue{..}

\maketitle

\section{Introduction} 
\label{Introduction}

Smart camera systems are changing the way consumers secure their homes and businesses. Commercial camera systems have been remarkably successful; they have become the \emph{de facto} monitoring system, as they offer the following essential services with plug-and-play support: (1) watch live and recorded video feeds, (2) share videos with others, (3) delete recorded videos, (4) recover access to the system, and (5) perform a full factory reset. Yet, while the market demand for
smart camera systems continues to grow rapidly as reported by \textit{Ring}~\cite{cameron_rings_2019, wyden_wyden_2019,herrman_whos_2020, huseman_huseman_2020}, \textit{Wyze}~\cite{wyze_wyze_2018}, and \textit{Arlo}~\cite{arlo_arlo_2021}, consumers have come to realize that the costs of owning a smart camera system are not exclusively monetary.

Commercially available smart camera systems follow a threat model that mandates undue trust \textit{by design}; the service provider is granted unfettered access to the video content of any consumer who uses their system. Ring has recently come under legal scrutiny~\cite{guariglia_lapd_2021,flores_bad_2020} for allowing more than 2,000 government agencies to directly request videos from users without formal due process~\cite{paul_amazons_2019, lecher_ring_2019, brewster_smart_2018,ring_active_2021}. Perhaps even more troubling, employees are viewing and annotating live user streams for research~\cite{deahl_ring_2019,biddle_for_2019} while others are abusing their access to view and share users' videos online~\cite{keck_amazons_2019, ropek_home_2021}. Moreover, research has posited that these systems can be transformed into mass surveillance systems, given their widespread adoption~\cite{cameron_rings_2019,ng_ring_2019,lecher_ring_2019,cameron_amazon_2019,bridges_amazons_2021,kraus_ring_2019,ring_ring_2019,ng_amazons_2019}.
The message behind the underlying design of modern smart camera systems is clear: users do not have control over their own videos and system, compromising user privacy.

To this end, we answer the following question: \textit{can users afford all of the features present in commercial smart camera systems, without compromising their privacy?} A cryptographic approach is a plausible way to protect users' privacy in that it enables users to solely assume control over videos stored in the cloud. However, practical realizations of such systems face several key challenges: cryptographic protections (e.g., encryption) incur computational overheads, affecting system performance, since stored videos are encrypted; a fine-grained sharing scheme (i.e., sharing specific video fragments) requires a user-controlled key management system; and generating, storing, rotating, and re-negotiating cryptographic keys poses further challenges on performance and usability. Troublingly, even if such challenges could be addressed, recent events have demonstrated that encryption alone is not sufficient to protect users from abuses by governments through coercion~\cite{kravets_indefinite_2016,greenberg_two_2021}. Thus, meeting performance, security, privacy, and usability goals demands a novel approach that is sensitive to the unique requirements of this domain.

In this paper, we present \cactus{}, a privacy-preserving smart \textbf{Ca}mera system \textbf{C}ontrolled \textbf{T}otally by \textbf{Us}ers. Inspired by information privacy laws, \cactus{} is designed to enforce three privacy goals through known security properties; (1) \emph{the right to not be seen}: the user is assured \textit{confidentiality} of stored videos and live video streams, (2) \emph{the right of sole ownership}: the user (and only the user) is trusted, and has \textit{complete mediation} over access to their data by others, (3) \emph{the right to be forgotten}: deleted videos are not recoverable, even in cases of coercion.

To meet the required feature set of commercial systems and address the stringent technical challenges and privacy goals of smart camera systems, we design \cactus{} as follows: it allows the user to solely assume control of the smart camera system through a direct and physical pairing process (that is, without relying on or trusting third parties); isolates and protects access to video footage through encryption, key rotation, and key management; enables viewing live and stored videos through performance-aware cryptographic algorithms; supports video deletion and factory reset via key rotation and management; and provides fine-grained (i.e., on the scale of seconds) peer-to-peer delegation of video footage through a binary key tree. We make the following contributions:

\begin{enumerate}

    \item We present \cactus{}, a privacy-preserving smart \textbf{Ca}mera system \textbf{C}ontrolled \textbf{T}otally by \textbf{Us}ers, that returns controls of the system to users without compromising features found in commercial smart camera systems.
    
    \item \ifrevision \hypertargetblue{mr1b}{\mr{1b}} \fi We perform a functional user evaluation of our system and find that \cactus{} is natural and easy to use, all while meeting our privacy goals. 
    
    \item We perform a performance evaluation of \cactus{} on a Raspberry Pi and find that we can serve a live video stream at a resolution of \resquality{}, at a frame rate of \resfps{}, and with a latency of \reslatency{}. 

\end{enumerate}

To encourage future privacy-preserving smart camera systems, we release \cactus{} as open-source software, available at \iffinal \url{https://github.com/siis/CaCTUs}. \else \texttt{ANONYMIZED FOR SUBMISSION}. \fi
\section{Background}
\label{Background}

\subsection{Smart Camera Systems}

A smart camera system is a collection of cameras that are connected to the Internet, allowing \textit{owners} (i.e., those who purchase and configure the system) to view live and recorded videos of their homes from anywhere.
Most companies sell their systems as an integrated ecosystem: cameras work with a purpose-built smartphone application that allows the owner to view footage, delegate access, and administer their smart camera system. At the core of these systems are five functions: (1) recording and streaming, (2) sharing (delegation), (3) deleting, (4) access recovery, and (5) factory reset. Each function places requirements and motivates the architecture of the ecosystems available to consumers.

\noindent
\textbf{Recording and Streaming}. Camera systems allow owners to view live and recorded footage from all cameras they own using an application on their smartphone, allowing them to monitor the current status of their property.
As the most fundamental function provided by camera systems, this is expected to work reliably and globally: users want to be able to view footage anywhere, and recover footage even in the case of physical failures of the camera or home Internet connection. To facilitate this, consumer smart camera systems currently entrust the data to a cloud provider, streaming camera data to cloud storage as it is captured and making it available to the owner's device. As a result, access to the footage is managed by the cloud provider, who must be trusted to prevent unauthorized access.

\noindent
\textbf{Delegation}. Owners want to share access to their camera systems with others. We refer to this capability as \textit{delegation}. Whether used to provide a house-sitter with access to live footage during a vacation, or sharing video of an incident after the fact, this delegation is expected to be \textit{fine-grained}, meaning it applies to specific users (\textit{delegatees}) for only the portion of time that they need access. Consumer smart camera systems allow policy enforcement as a means of delegation: each user has an attached policy for the time range of live or recorded footage they may access, and cloud storage mediates this to prevent delegatees from exceeding their policies.

\begin{figure}[t]
    \centering
    \includegraphics[width=\columnwidth]{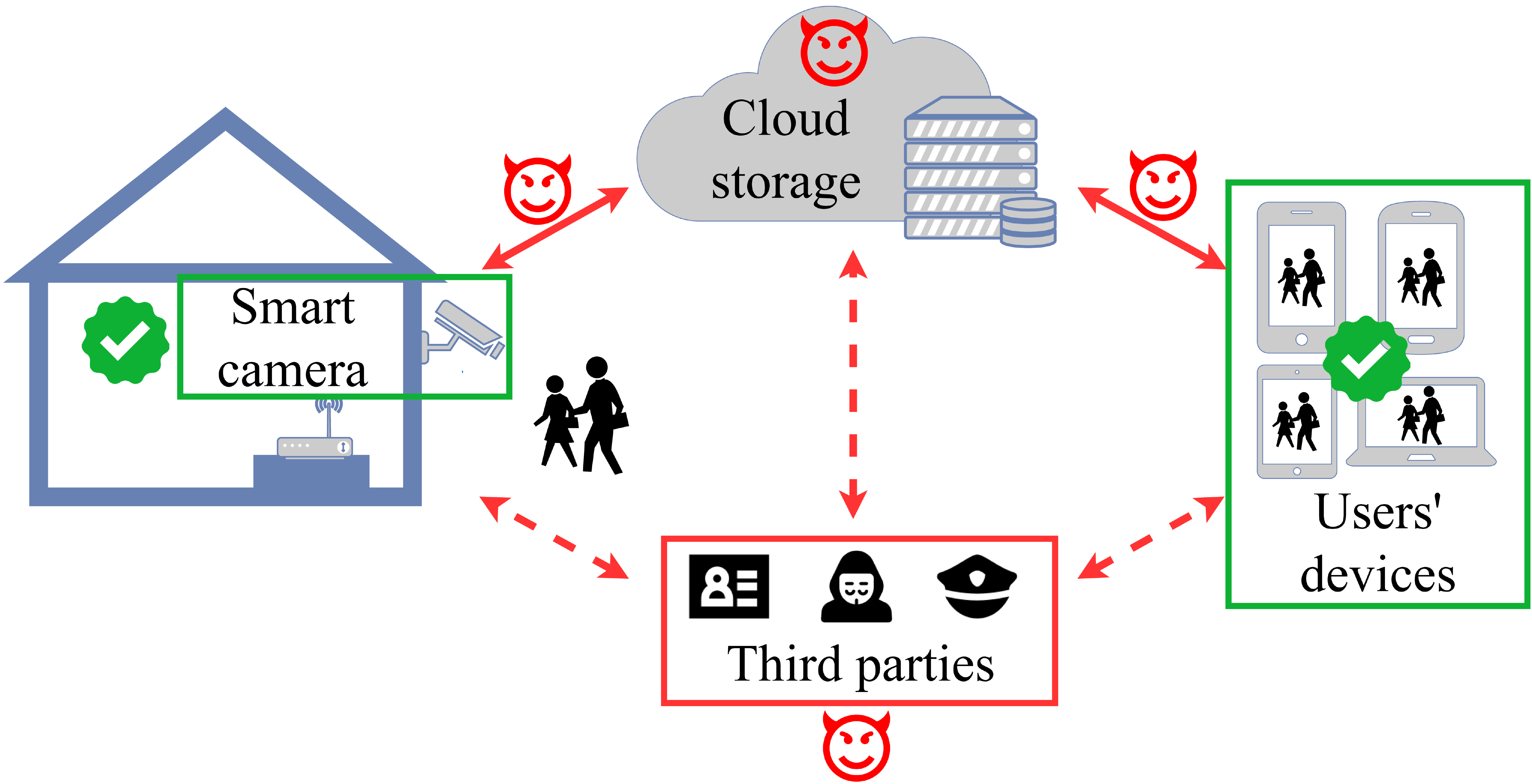}
    \caption{Overview of the components and actors in the privacy-preserving smart camera system \cactus{}: the camera devices and the users' devices are trusted while the cloud storage provider, the networks, and any other third party are untrusted.}
    \label{fig:threatmodel}
\end{figure}

\noindent
\textbf{Deletion}. Owners expect the ability to fully delete their data to prevent further access by any party. The right of consumers to delete their personal data has been codified in legal frameworks, such as the European General Data Protection Regulation (GDPR)~\cite{european_parliament_regulation_2016} and California Consumer Privacy Act (CCPA)~\cite{california_state_legislature_title_2018}. As recorded data from camera systems are saved to the cloud, owners must trust cloud providers to delete their data when requested, including copies stored elsewhere in the cloud.

\begin{figure*}[!ht]
    \centering
    \includegraphics[width=.9\textwidth]{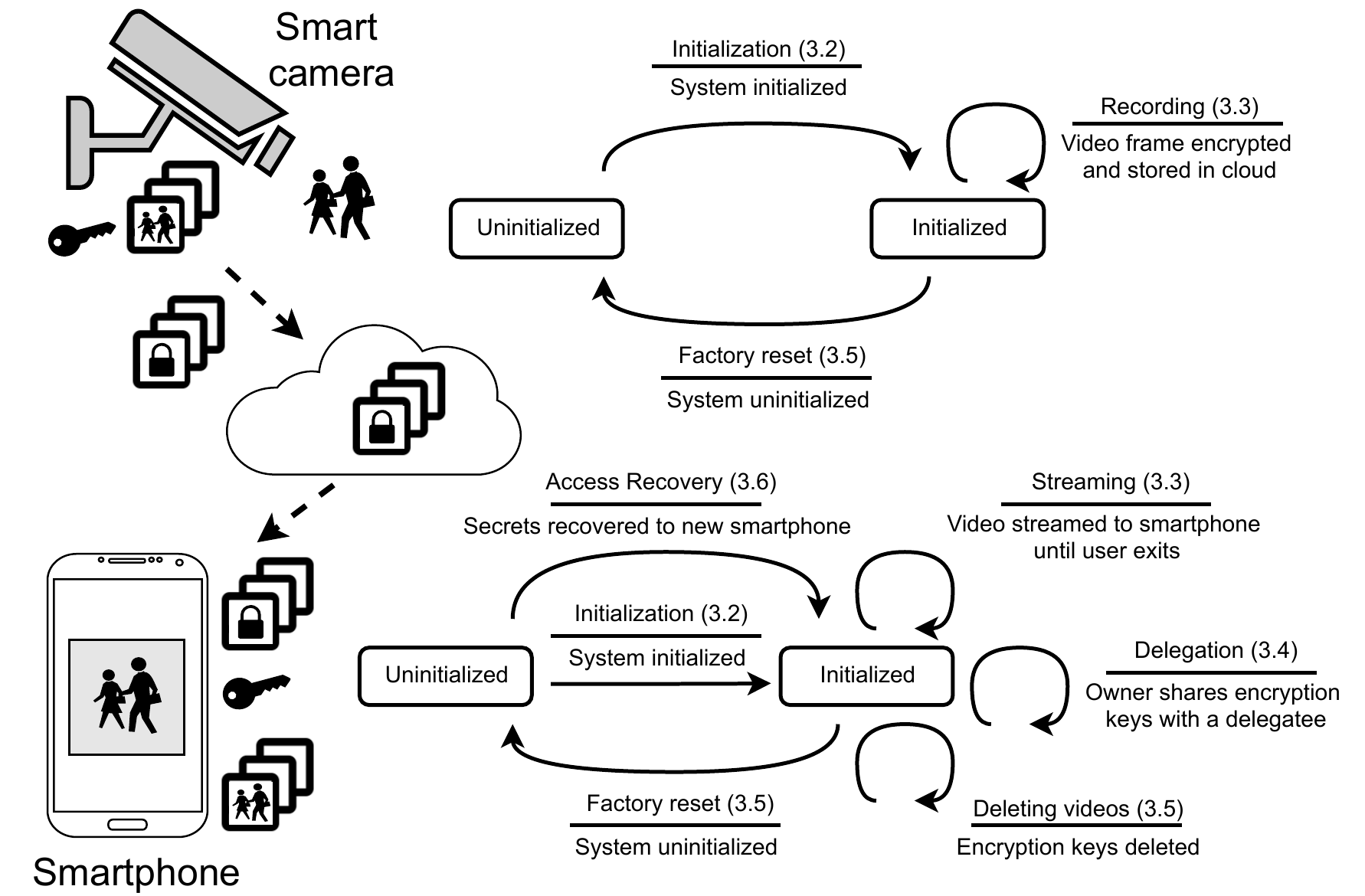}
    \caption{Overview of the recording and streaming operations; the camera records video frames, encrypts them locally, sends the encrypted frames to the cloud, from where the smartphone application downloads them, decrypts them locally, and plays the video. The right part is the overview of the states and actions in the \cactus{} system for the camera at the top right and the owner's smartphone at the bottom right (references indicated between parentheses are the subsections to refer to for more details).}
    \label{fig:overview-cactus}
\end{figure*}

\noindent
\textbf{Access Recovery}. Since access to smart camera systems is mediated by a set of credentials (e.g., a username and password), these systems must account for the possibility of a user losing these credentials. When authentication is performed by a cloud service, this is relatively straightforward: the user's identity is verified via other means, such as a password reset through email. As we will discuss, however, such recovery is only trivial because of the trust assumptions of these systems, and we will see that this critical function requires careful thought under other trust models.

\noindent
\textbf{Reset}. Finally, owners may wish to stop use of the smart camera system. In this case, they will expect all of their stored footage to be deleted, access to live footage revoked, and the device returned to a condition where it may be set up by another user. This is generally equivalent to a delete operation for all stored data, followed by resetting the physical camera itself.

\subsection{Privacy in Smart Camera Systems}
\label{privacy}

Smart camera systems have been shown to have both privacy and security risks~\cite{paul_amazons_2019, lecher_ring_2019, brewster_smart_2018,ring_active_2021,deahl_ring_2019,biddle_for_2019, keck_amazons_2019,ropek_home_2021,cimpanu_hackers_2019}. As a result, information privacy laws have been passed, which aim to address expectations of privacy in online and physical environments (e.g., the California Consumer Privacy Act (CCPA), California Privacy Rights Act (CPRA), and the European General Data Protection Regulation (GDPR), among others~\cite{california_state_legislature_title_2018,  california_state_legislature_california_2020, european_parliament_eprivacy_2009, european_parliament_regulation_2016, house_of_commons_of_canada_bill_2020, the_constitution_project_guidelines_2007, european_data_protection_board_guidelines_2019}). However, these legal frameworks often suggest vague, rather than concrete requirements for enforcing privacy in specific end-user devices such as smart cameras. Motivated by these recommendations and requirements necessary to prevent previously-discussed privacy incidents, we can achieve a \textit{privacy-preserving system} by affording the system owner the following rights \ifreviews \hypertargetblue{mr3a}{\mr{3a}} \fi:

\begin{enumerate}
    \item \textbf{Right to not be seen}: the owner is assured that no unauthorized user can view stored videos or live video streams.
    \item \textbf{Right of sole ownership}: the owner retains full control of their data and who they trust.
    \item \textbf{Right to be forgotten}: deleted videos are not recoverable, even in cases of coercion.
\end{enumerate}

In practice, these rights imply that device owners must have exclusive control over the collection of data, its uses, and the access delegations to it. 

\subsection{Threat Model}
\label{ThreatModel}

Our goal in this work is to demonstrate a smart camera system that provides feature parity with commercial systems while placing no trust in a cloud provider or other third party. As such, we work under a threat model wherein edge devices (i.e., the smart camera and end-user devices) are trusted, but the cloud storage provider, network, and any other third-party service are untrusted (see \autoref{fig:threatmodel} for an overview of \cactus{}).

We only trust the devices owned by the users (cameras, smartphones, laptops, or tablets) to securely handle the encryption and decryption keys used in the system, and we trust the device manufacturer to provide us with a camera device that correctly executes its functionality. This additionally implies that supply-chain exploits against the camera manufacturer are out of scope. We trust the other applications running on the users' devices (or that the operating system sufficiently isolates these applications) and we assume that the cryptographic algorithms used provide the advertised guarantees (e.g., Diffie-Hellman assumption~\cite{diffie_new_1976} and RSA public-key cryptosystem~\cite{rivest_cryptographic_1983}). \ifreviews \hypertargetblue{rb7}{\re[7]{B} (secure timing deleted)} \fi

We also acknowledge that access under our system may be universally delegated: once granted access to a video, a party is not prevented from sharing with others or downloading and storing the videos somewhere else. Partial mitigations to this may be considered, but as such sharing can occur outside the purview of our system complete prevention is not possible. \ifrevision \hypertargetblue{mr4a}{\mr{4a}} \fi Encrypted frames are assumed to be publicly accessible (as we do not trust the cloud to do any access control mediation), thus we acknowledge that access pattern to the cloud storage may be leaked in \cactus{}. Ongoing research in Private Information Retrieval (PIR) or Oblivious RAM (ORAM) could provide potential mitigations through the use for example of random accesses and dummy writes to the cloud storage. Finally, \ifreviews \hypertargetblue{ra3}{\re[3]{A}} \fi physically tampering with the devices (modifying the hardware, chip-level changes to edit the software execution, etc.) and denial of service attacks are outside the scope of our work.
\section{CaCTUs}
\label{CaCTUs}

\subsection{Overview}
\label{cactus_overview}

 \ifreviews \hypertargetblue{mr3b}{\mr{3b}} \fi In the following sections, we will describe how \cactus{} meets the privacy goals described in \autoref{Background} by providing the following three security properties: (1) \emph{confidentiality}: stored videos and live video streams cannot be viewed by unauthorized parties, (2) \emph{complete mediation}: the user fully controls access to their data by others, and (3) \emph{deletion}: deleted videos are not recoverable, even in cases of coercion. Each subsection motivates the privacy goals before providing technical details of the feature. We refer to \autoref{Appendix:notation} for the notation used in our cryptographic constructions and algorithms, to \autoref{Appendix:protocol} for the details of the protocols, and to \autoref{Appendix:storyboard} for the storyboard of the \cactus{}'s smartphone application. For clarity, we consider only a single camera, though our approach readily generalizes to multi-camera systems by applying the described protocols to each camera individually.

\autoref{fig:overview-cactus} shows an overview and state diagram of \cactus{}. The camera locally encrypts recorded video frames before uploading them to cloud storage. The smartphone application performs the reverse operations: it downloads the frames, decrypts them locally, and plays the video. Regular key rotation and secure key management allow the system to support secure streaming, delegation, deletion, recovery, and reset.

\begin{table}[t]
\centering
\begin{tabular}[t]{m{0mm}cccc}
\hline
 & Camera & \faEye & \faBluetoothB & Smartphone\\ \hline
 
\multirow{3}{*}{1} & \multirow{3}{*}{\includegraphics[height=3\baselineskip]{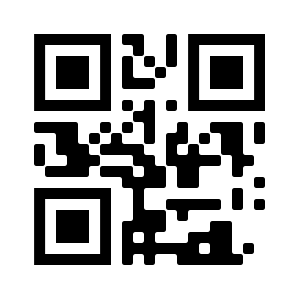}} & \multirow{3}{*}{$\overset{ h_{PK_f}}{\longrightarrow}$} & &\\ 
 &&&&\\
 &&&&\\

\multirow{4}{*}{2} & & & \multirow{4}{*}{$\overset{ PK_f}{\longrightarrow}$} & \multirow{2}{*}{$h'_f = hash(PK_f)$}  \\ 
 & & & & \multirow{2}{*}{$h'_f\overset{?}{=} h_{PK_f}$} \\
&&&&\\
&&&&\\

 \multirow{2}{*}{3} & \multirow{2}{*}{$h'_o= hash(PK_o)$} & &  \multirow{2}{*}{$\overset{ PK_o}{\longleftarrow}$}  &  \multirow{2}{*}{$gen(SK_{o}, PK_{o})$}\\
 &&&&\\
 
 \multirow{3}{*}{4} & \multirow{3}{*}{$h'_o \overset{?}{=} h_{PK_o} $} & \multirow{3}{*}{$\overset{ h_{PK_o}}{\longleftarrow}$} & &\multirow{3}{*}{\includegraphics[height=3\baselineskip]{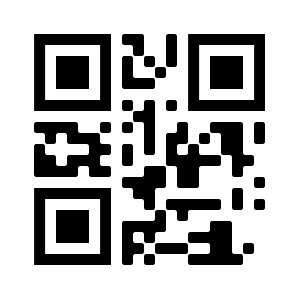}} \\ 
 &&&&\\
 &&&&\\

  \multirow{2}{*}{5} & \multirow{2}{*}{$DH(SK_{f},PK_{o})$} & \ifreviews {\tiny \hypertargetblue{ra2b}{\re[2b]{A}}} \fi &  \multirow{2}{*}{$\overset{verif}{\longleftrightarrow}$} & \multirow{2}{*}{$DH(SK_{o}, PK_{f})$}\\ 
 &&&&\\
 
 \multirow{3}{*}{6} & \multirow{3}{*}{$gen(SK_{c}, PK_{c})$} & \ifreviews {\tiny \hypertargetblue{ra2a}{\re[2a]{A}}} \fi & \multirow{3}{*}{$\underset{(RSA)}{\overset{PK_{c}}{\longrightarrow}}$} &\\
  &&&&\\
  &&&&\\
  
 \multirow{2}{*}{7} & \multirow{2}{*}{$init(secrets)$} & \ifreviews {\tiny \hypertargetblue{ra2a}{\re[2a]{A}} }\fi & \multirow{2}{*}{$\underset{(RSA)}{\overset{secrets}{\longleftarrow}}$}
 & \multirow{1}{*}{$gen(escrow)$}\\ 
 & & & & \multirow{1}{*}{$passphrase$} \\
 
\hline
\end{tabular}
\caption{Protocol followed by the camera and the owner's smartphone during the initialization, \faEye{} and \faBluetoothB{} respectively correspond to what is obtained through the visual and Bluetooth channels.}
\label{tab:initialization}
\end{table}

\subsection{Initialization}
\label{initialization}

During initialization, the user's smartphone and camera establish a trust association used for all other steps, so the security of this step is critical to that of the system as a whole.  In \cactus{}, we adapt the \textit{Seeing-Is-Believing} (SiB) technique introduced by McCune, Perrig, and Reiter~\cite{mccune_seeing-is-believing_2005} to establish an authenticated communication channel between the camera and smartphone when the devices share no prior context (and without having to trust any third party). Specifically, we use the visual channel as an out-of-band means to verify the authenticity of each end of a Bluetooth channel (see \autoref{fig:setup_system}). 

\ifreviews \hypertargetblue{mr3c}{\mr{3c}} \fi This secure pairing bootstraps the system to allow the owner of the device to communicate directly with the camera (for system initialization) while ensuring confidentiality and integrity. Thus, negotiation of encryption keys can be done without any other party involved. Moreover, the required proximity and physical interaction between the devices would render attack attempts easily detectable by the system owners. The initialization protocol between the camera and the owner's smartphone application is as follows (see \autoref{tab:initialization}):

\begin{figure}[t]
    \centering
    \includegraphics[width=0.49\columnwidth]{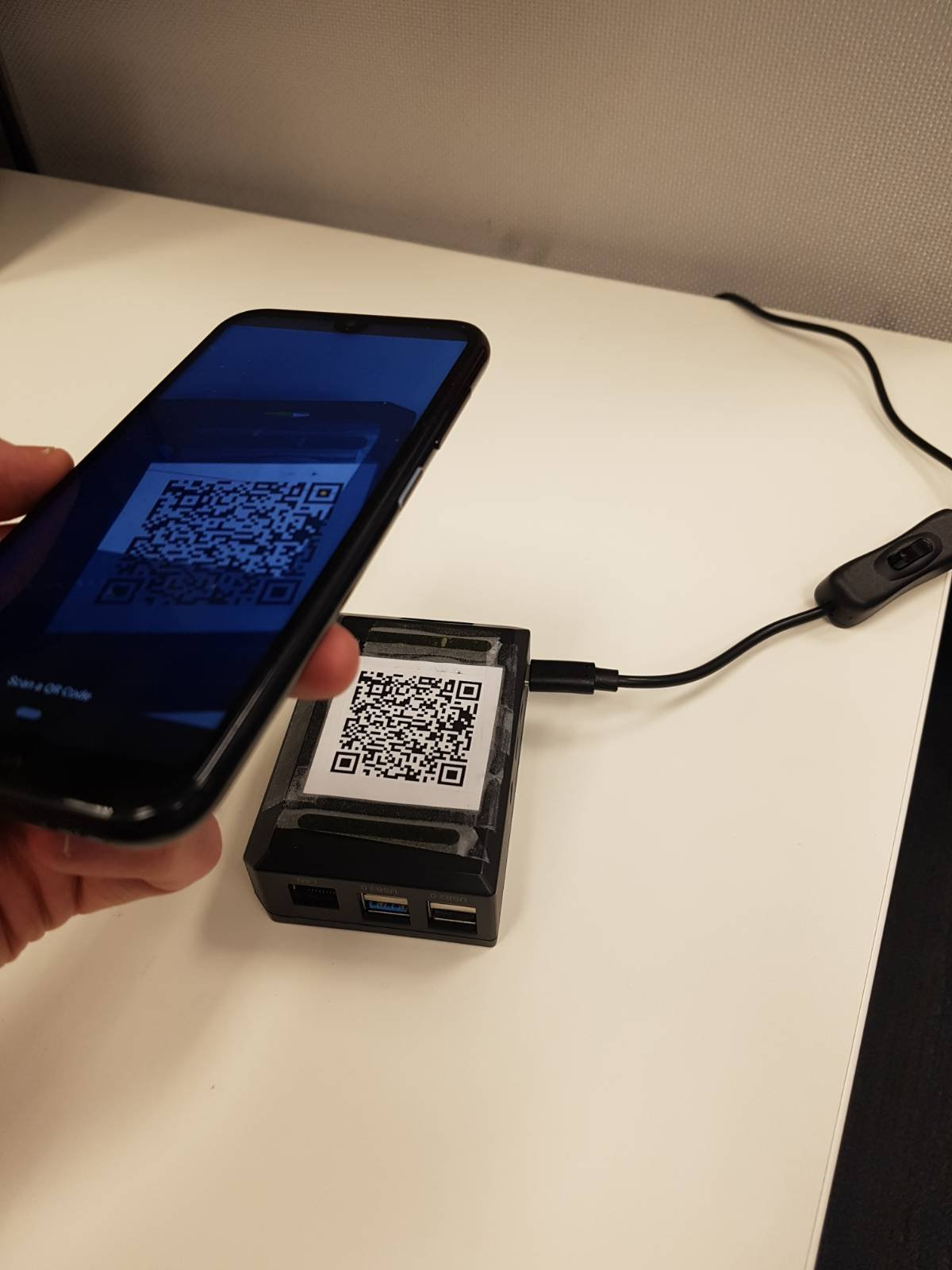}
    \includegraphics[width=0.49\columnwidth]{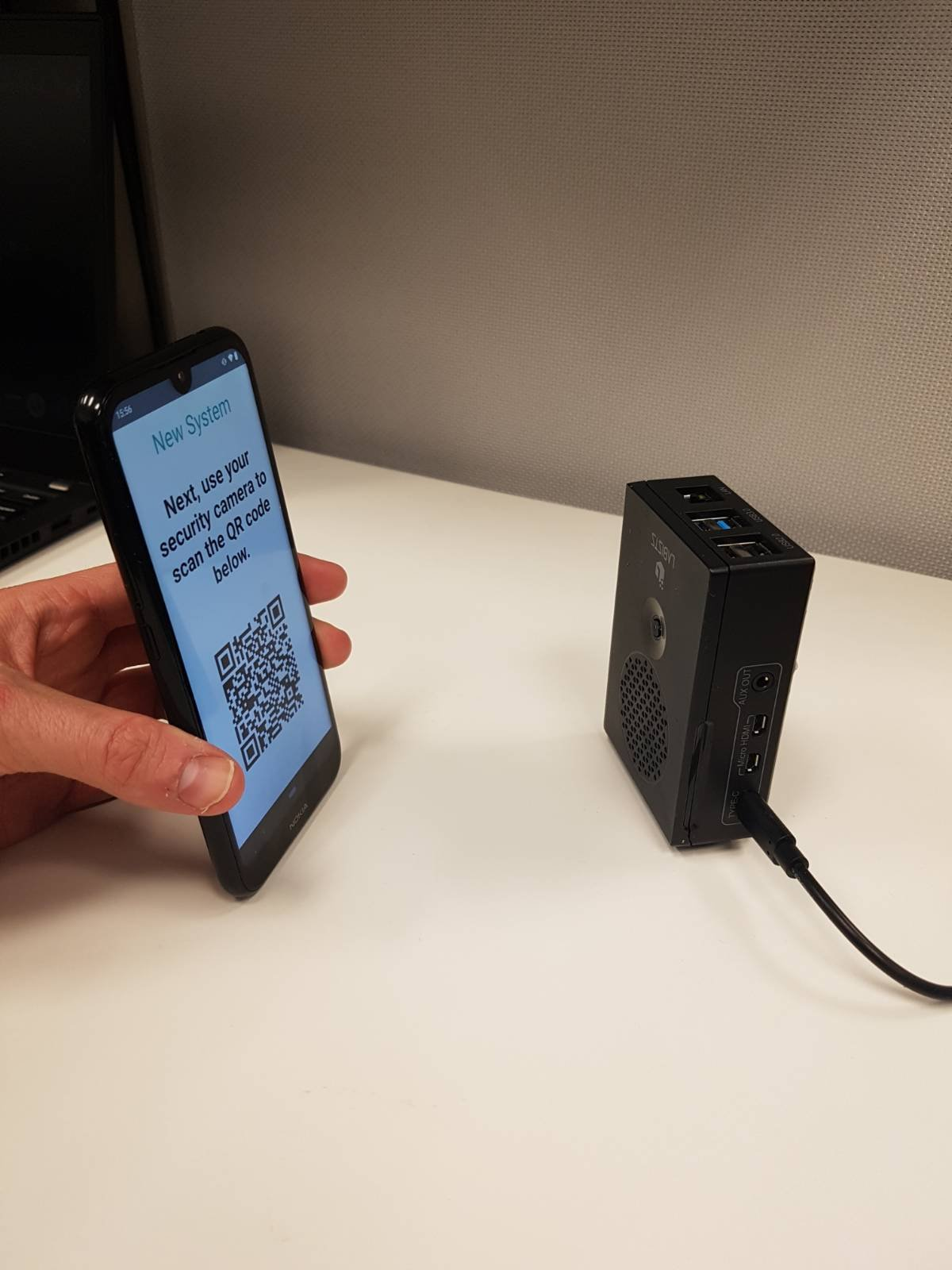}
    \caption{Establishment of a secure and authenticated Bluetooth pairing to setup \cactus{} by scanning QR codes (visual channel).}
    \label{fig:setup_system}
\end{figure}

\begin{enumerate}
    \item A pair of factory-generated asymmetric keys $(SK_{f}, PK_{f})$ is present on the camera device, and the hash of its public key $h_{PK_f}$ is embedded into a QR code on the back of the device (recall from \autoref{ThreatModel} that the supply chain is trusted). The owner's smartphone scans this QR code and stores its content.
    
    \item The camera and the owner's smartphone connect through Bluetooth and the camera sends its public key $PK_{f}$ to the owner, who computes the hash of the camera's public key $h'_f$ and checks that it matches the hash retrieved from the QR code. 
    
    \item If they match, the owner's smartphone generates its own asymmetric pair of keys $(SK_{o}, PK_{o})$ and sends its public key $PK_{o}$ through Bluetooth, the camera computes the hash $h'_o$ of the owner's public key.
    
    \item The owner points their smartphone's screen at the camera. On the screen is displayed the QR code with the hash of the owner's public key. The camera retrieves the content from the QR code and checks that it matches $h'_o$.
    
    \item If the key hashes match, \ifreviews \hypertargetblue{ra2b}{\re[2b]{A}} \fi both devices now verify that the other device knows the secret key corresponding to the public key that they advertised earlier. This is can be done for instance by applying the Diffie-Hellman key exchange to compute their shared secret and then by exchanging a series of encrypted messages where both parties prove their knowledge.

    \item The camera then generates a new asymmetric key pair $(SK_{c}, PK_{c})$ that it will use for future communication (to avoid further reliance on the factory-generated key). \ifreviews \hypertargetblue{ra2a}{\re[2a]{A}} \fi The camera then shares its public key $PK_{c}$ with the owner through Bluetooth in an authenticated way using RSA and the factory-generated asymmetric key pair $(SK_{f}, PK_{f})$. \ifrevision \hypertargetblue{mr3d}{\mr{3d}} \fi Note that we could have used the shared secret computed at the previous step through Diffie-Hellman, however, we chose to use RSA to align this step of our initialization protocol with the future protocols presented in \cactus{} as well as because we want authentication of the messages.
    
    \item Next, the owner sends their \textit{secrets} to the camera device to complete system setup, \ifreviews \hypertargetblue{ra2a}{\re[2a]{A}} \fi this is done in a secure and authenticated way using RSA, but this time the new asymmetric key pair of the camera is used $(SK_{c}, PK_{c})$. First, they send \textit{wifi credentials} of the wifi network the camera should connect to. Then, they generate and send a \textit{seed key} that will be used to derive the keys to encrypt video frames. Lastly, they send \textit{escrow material} (protected by a non-recoverable \textit{passphrase}) that may be used by the owner to recover access to the system (see \autoref{recovery} for details about the escrow material). \ifrevision \hypertargetblue{mr3e}{\mr{3e}} \fi If necessary, during this step the owner could also configure the cloud storage option they want to use if different than the default one.

\end{enumerate}

At this point, the system is initialized, the camera begins recording, and the owner can begin executing the other camera functions.

\subsection{Recording and Streaming Videos}
\label{viewing}

\ifreviews \hypertargetblue{mr3d}{\mr{3d}} \fi Here we describe how to ensure the \textit{confidentiality}, \textit{integrity}, \textit{authenticity}, and \textit{freshness} of the recorded video footage. The video frames and metadata are encrypted locally at the camera and asymmetrically signed in blocks of $N$ frames, before being uploaded to cloud storage. \ifreviews \hypertargetblue{rb2a}{\re[2a]{B}} \fi At each key rotation, the camera device securely erases the encryption keys previously used as it does not need them anymore. To view a video, users of the system download the encrypted data from the cloud storage, derive the decryption keys locally (if they have access to them), and decrypt the frames to rebuild the video. Thus, only users with access to the appropriate decryption keys can view the footage. We discuss in \autoref{Appendix:signing} how to leverage one-time signatures (i.e., hashed signatures \cite{gennaro_how_2001}), to build a scheme that allows us to sign every frame\footnote{The overhead of asymmetrically signing every frame is sufficiently large that it decreases the performance of the system by several orders of magnitude; thus the need for hashed signatures.}. Nonetheless, as the camera is recording, it performs the following:

\begin{enumerate}
    \item Consider a block of $N$ frames. Each frame $F_i$ is recorded at timestamp $t_i$.
    \item A key rotation scheme is used to derive encryption keys for a given frame (this will prove useful for delegation, discussed in \autoref{delegation}). The rotation scheme $\mathcal{K}$ provides a key $k_i$, the key used to encrypt frame $F_i$. An initialization vector $IV_{i}$ is randomly generated. 
    \[
    \{ k_{i} = Extract(\mathcal{K}, i) | i\in \llbracket 1, N \rrbracket\}
    \]
    \[
    \{ IV_{i} = RandBytes(16) | i\in \llbracket 1, N \rrbracket\}
    \]
    \item Next, each frame is symmetrically encrypted (\textbf{confidentiality)} into the corresponding ciphertext $C_i$ using the AES algorithm in Galois/Counter Mode (GCM, chosen for its performance) with a 256-bit key. $C_i$ is then concatenated with $IV_{i}$ and $t_i$ to be hashed into $h_i$ (\textbf{integrity and freshness}).
    \[
    \{ C_i = AES256Enc(IV_{i}, k_{i}, F_i) | i\in \llbracket 1, N \rrbracket\}
    \]
    \[
    \{h_i = HMAC(k_{i}, C_i||IV_{i}||t_i) | i\in \llbracket 1, N \rrbracket\}
    \]
    \item A signature $\sigma$ of the block is computed using the private key $SK_{c}$ of the camera and the $N$ hashes of the frames in this block (\textbf{integrity and authenticity}).
    \[
    \sigma = Sign(SK_{c}, h_1||h_{2}||...||h_{N})
    \]
    \item The encrypted and authenticated frames \ifreviews \hypertargetblue{ra2c}{\re[2c]{A}} \fi$\left< \{C_i,IV_{i},t_i | i\in \llbracket 1, N \rrbracket\}, \sigma \right>$ are uploaded to the cloud along with their corresponding metadata (initialization vector used for encryption and timestamp).
\end{enumerate}

Each user who has access to the correct decryption keys can download these encrypted and authenticated frames $\left< \{C_i,IV_{i},t_i | i\in \llbracket 1, N \rrbracket\}, \sigma \right>$. To view the video, the user performs the following:

\begin{enumerate}
    \item We consider a block of $N$ frames of signature $\sigma$ downloaded on demand. Each ciphertext $C_i$, encrypted using the initialization vector $IV_{i}$, corresponds to a frame recorded at timestamp $t_i$ .
    
    \item The corresponding symmetric key $k_{i}$ is extracted from key rotation scheme $\mathcal{K}$. The hash $h_i$ of each ciphertext $C_i$ is computed.
    \[
    \{ k_{i} = Extract(\mathcal{K}, i) | i\in \llbracket 1, N \rrbracket\}
    \]
    \[
    \{ h_i = HMAC(k_{i}, C_i||IV_{i}||t_i) | i\in \llbracket 1, N \rrbracket\}
    \]
    \item The signature $\sigma$ of the block is verified with the public key $PK_{c}$ of the camera and the $N$ hashes of the frames in this block (\textbf{authenticity, integrity, and freshness}).
    \[
    1 \overset{?}{=} Verify(PK_{c}, \sigma, h_1||h_{2}||...||h_{N})
    \]
    
    \item If the signature is correct, each ciphertext $C_i$ is then symmetrically decrypted into the corresponding frame $F_i$ (\textbf{confidentiality}).
    \[
    \{ F_i = AES256Dec(IV_{i}, k_{i}, C_i) | i\in \llbracket 1, N \rrbracket\}
    \]
\end{enumerate}

\ifreviews \hypertargetblue{mr3e}{\mr{3e}} \fi Encryption is performed \textit{end-to-end}: data is encrypted locally at the camera before being stored in the cloud and decrypted locally at the smartphone after being retrieved. Furthermore, integrity is ensured using an authenticated encryption scheme, so that the video footage cannot be tampered with during transmission or storage. Lastly, the identity of the camera is embedded into the video frames (and signed) so that users can attest the authenticity of the video footage. This scheme allows the user to verify the integrity, authenticity, and freshness of an arbitrary set of video frames, so that they then decrypt the frames and rebuild the video for playback. 

\begin{figure}[t]
    \centering
    \includegraphics[width=\columnwidth]{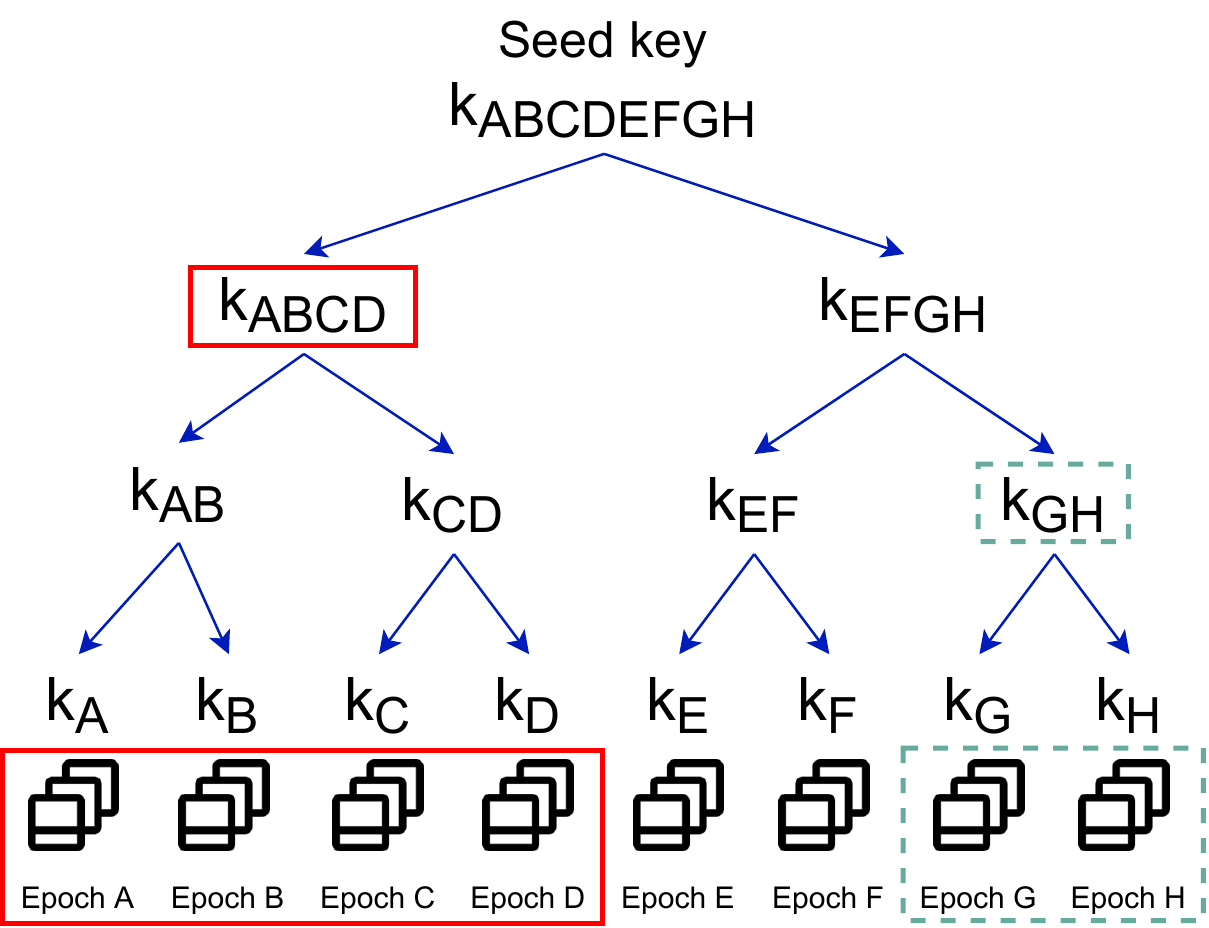}
    \caption{Key Tree Construction used by \cactus{}, the keys to be shared to give access to the corresponding footage are framed.}
    \label{fig:key_tree}
\end{figure}

\subsection{Delegation}
\label{delegation}

\ifreviews \hypertargetblue{mr3f}{\mr{3f}} \fi To protect privacy, users must have complete mediation over access to their videos. \cactus{} achieves this by ensuring that the owner has control of the keys used to encrypt the video footage. However, they may also want to delegate access to their videos (e.g., to friends or family) for different periods of time. Achieving fine-grained sharing capabilities for delegatees is nontrivial: we want to support delegation without knowing beforehand to whom the owners will delegate access or for how long. To enable this, we rotate the key used to encrypt the video frames at the end of every \textit{epoch} (a fixed-size time interval). We use a binary key tree construction to facilitate the management of all the keys for the camera device, owners, and delegatees. A peer-to-peer pairing is adopted for sharing keys so that there is no reliance on a third party.

Recall that frames are encrypted using a symmetric key derived from a key rotation scheme $\mathcal{K}$. In practice, to support delegation, this rotation scheme is a binary key tree, inspired by the key tree introduced by Kocher~\cite{kocher_complexity_2011}. The tree is of a fixed depth $d_\mathcal{K}$. In the tree, each leaf node holds some cryptographic key $k$ and covers a specific epoch: a time interval $[t_j, t_{j+1})$ of fixed-size $\delta_\mathcal{K}$. The root node of the tree is initialized with a seed key that is negotiated during the initialization of \cactus{} (see \autoref{initialization}). The timestamp of this negotiation is used for $t_0$, with $t_{j+1}=t_j+\delta_\mathcal{K}$. The leaf nodes in the tree hold the encryption keys for every epoch. Each node in the tree can be derived from the root node knowing the derivation equations and relations between the parent node and its two children.

\begin{figure}[t]
    \centering
    \includegraphics[width=0.6\columnwidth]{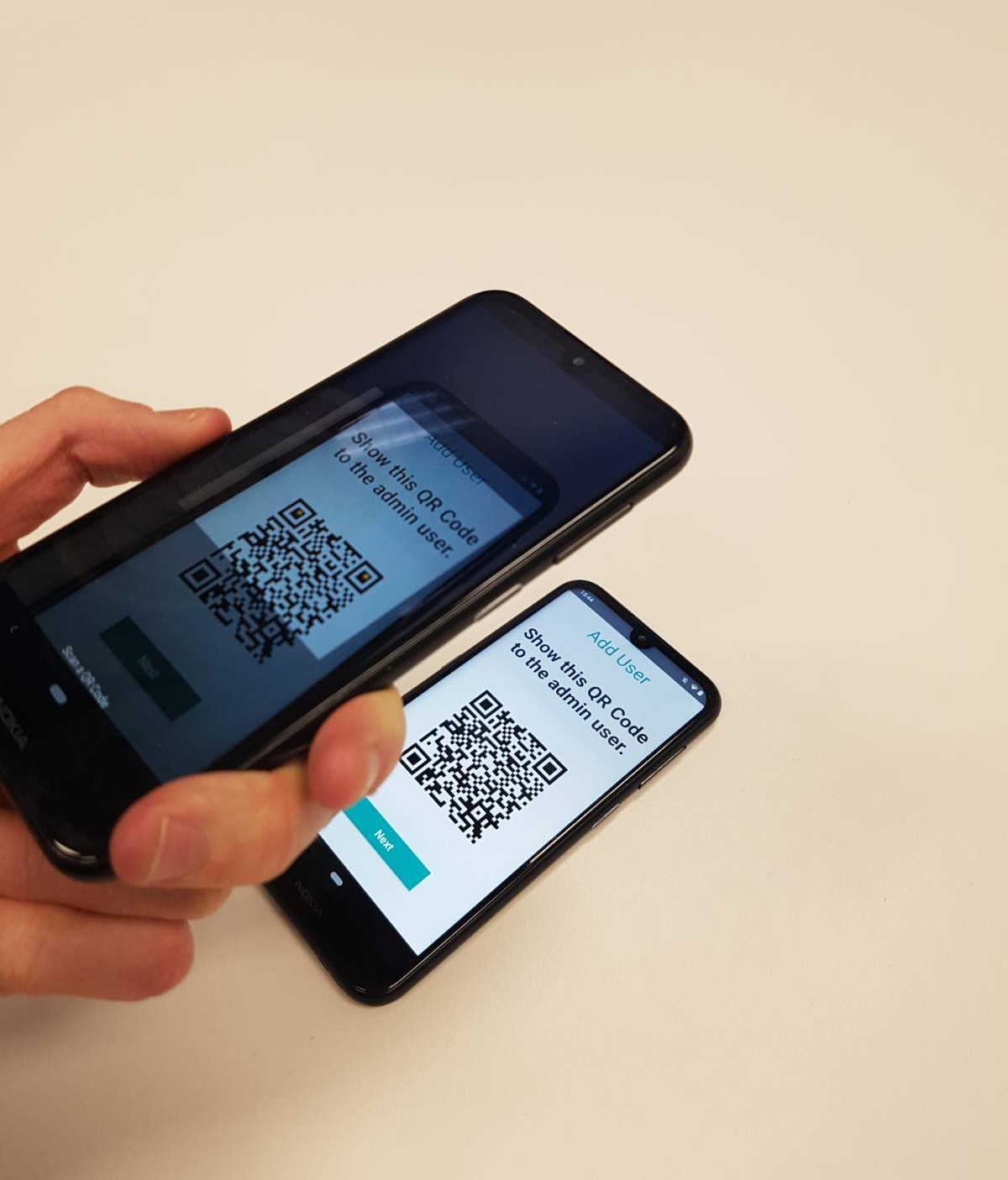}
    \caption{The establishment of a secure and authenticated Bluetooth pairing with a shared user's device to delegate access.}
    \label{fig:deleg}
\end{figure}

Within each epoch, the symmetric key used for each frame is identical. The derivation of keys is based on a Hash-Based Key Derivation function (HKDF), which is a one-way process \cite{katz_introduction_2014}. If we have $k_{parent}$, then:
\begin{equation}
\begin{gathered}
k{_{left}} = HKDF(k_{parent})\\
k{_{right}} = HKDF(k_{parent} \oplus{} 1)\\
\end{gathered}
\end{equation}
\ifreviews \hypertargetblue{mr3g}{\mr{3g}} \fi Therefore, for a given key in the tree, a user can only derive the keys below it but not the ones above (see \autoref{fig:key_tree} for an illustration). The binary key tree is useful for several reasons. First, it decreases the amount of keys that need to be shared with the delegatees, as the derivation algorithm is publicly known by the Kerckhoffs's principle and a specific part of the key tree can be reconstructed on-demand by a delegatee that is given access to a node of the tree. Moreover, it is storage space-efficient, as the key rotation mechanism generates a large number of encryption keys. \ifreviews \hypertargetblue{rb2b}{\re[2b]{B}}\fi When key rotation happens and the camera device securely erases the previous encryption keys, the camera device only needs to save at most $d_\mathcal{K}$ nodes to keep functioning. Further, it is simple to find which node is responsible for the encryption or decryption of a specific timestamped frame (through a binary search). \ifreviews\hypertargetblue{rb2c}{\re[2c]{B}}\fi Lastly, it facilitates the recovery process for the owners as only a subset of keys needs to be recovered to rebuild the tree from the escrow material.

Delegation is supported by giving some keys of the key tree to each delegatee (an example is given in \autoref{fig:key_tree} where keys to share and their corresponding epochs are framed). From there, each delegatee can derive the associated keys to correctly decrypt and view only video footage captured within the time window they were authorized for. \ifreviews \hypertargetblue{ra4a}{\re[4a]{A}} \fi The depth $d_{\mathcal{K}}$ and the epoch size $\delta_{\mathcal{K}}$ are configurable parameters of the system. For our implementation, we selected $d_{\mathcal{K}} = \SI{32}{}$ and $\delta_{\mathcal{K}} = \SI{10}{s}$. Importantly, this choice of parameters demonstrates the worst-case performance users can expect from \cactus{}. Specifically, these parameters result in a key tree that covers a lifespan of \textit{\SI{1362}{years}} at a \SI{10}{s} level of granularity. We provide more reasonable parameter choices (and, thus, expected performance at deployment) in \autoref{tab:deletion_space}. The size of an epoch $\delta_\mathcal{K}$ corresponds to the lowest delegation granularity achievable in \cactus{}. Delegation uses a peer-to-peer approach similar to the secure and authenticated pairing presented in \autoref{initialization}. The only difference being that there is no need to use a factory-generated key and QR code as both devices have a screen and can therefore generate their initial asymmetric key pairs dynamically. Exact details of this protocol can be found in \autoref{Appendix:delegation}.

\begin{table}[ht]
\centering
\begin{tabular}{cccc}
\hline
& \multicolumn{2}{c}{Lifespan for}  & \multicolumn{1}{c}{Storage space}\\
$ d_{\mathcal{K}} $ & $\delta_{\mathcal{K}} = \SI{10}{s}$ & $\delta_{\mathcal{K}} = \SI{60}{s}$ & (worst-case scenario) \\ \hline
\SI{24}{}    & \SI{5}{years}          & \SI{32}{years}          & \SI{256}{MB}           \\
\SI{26}{}    & \SI{21}{years}          & \SI{128}{years}         & \SI{1}{GB}             \\ 
\SI{28}{}    & \SI{85}{years}          & \SI{511}{years}         & \SI{4}{GB}              \\ 
\SI{30}{}    & \SI{340}{years}         & \SI{2043}{years}        & \SI{16}{GB}             \\ 
\SI{32}{}    & \SI{1362}{years}        & \SI{8172}{years}        & \SI{64}{GB}            \\ 
\end{tabular}
\caption{Lifespan of the key tree for an epoch time $\delta_{\mathcal{K}}$ of \SI{10}{s} and \SI{60}{s} and storage space needed to save to disk the encryption keys in the worst-case scenario for different depth sizes $d_{\mathcal{K}}$.}
\label{tab:deletion_space}
\end{table}

\vspace{-3\baselineskip}
\subsection{Deleting Videos and Factory Reset}
\label{deletion}

\ifreviews \hypertargetblue{mr3h}{\mr{3h}} \fi \cactus{} users must be able to delete their videos and factory reset their system. Owners can achieve this by deleting select decryption keys (i.e., a subset of nodes in the key tree) so that the keys below them in the tree cannot be recomputed. \ifreviews \hypertargetblue{rb2d}{\re[2d]{B}} \fi Recall that the camera device is securely deleting the encryption keys as soon as it does not need them anymore due to key rotation. As the keys used to encrypt the video frames are at the leaves of the key tree, to prevent being able to recompute such leaves, each node (i.e., key) along the path to the leaf must also be deleted. \ifreviews \hypertargetblue{ra4b}{\re[4b]{A}} \fi Thus, with a tree depth of $d_\mathcal{K}$, for each portion of video content composed of $n_{e}$ epochs to delete, the upper bound of the number of nodes that must be deleted from the tree is $\mathcal{O}(d_\mathcal{K}n_{e})$.

To provide an example: in \autoref{fig:key_tree}, to delete $k_A$ (the key for epoch A), keys $\{k_{AB}$, $k_{ABCD}$, $k_{ABCDEFGH}\}$ must also be deleted. Thus, $\{k_B, k_{CD}, k_{EFGH}\}$ must be saved in this sparser key tree so that the corresponding videos can still be decrypted. \ifreviews \hypertargetblue{rb2e}{\re[2e]{B}} \fi The key material in the escrow material also needs to be updated accordingly. \ifreviews \hypertargetblue{rb5}{\re[5]{B}} \fi\autoref{tab:deletion_space} presents the storage space required in the worst-case scenario where every other epoch has been deleted. Note that in practice, such a scenario is very unlikely to happen as it will render the system unusable. Furthermore, the owner is very unlikely to delete beforehand keys that would have been used to encrypt future videos. Thus, in practice far less amount of storage space is required, specifically thanks to our binary key tree structure that enables dynamic derivation of lower keys.  In this way, the binary key tree allows the owner to delete video footage at arbitrary time scales. Note that this operation is indeed equivalent to deleting the video frames as they can not be correctly decrypted without the keys, even in cases of coercion.

\ifreviews \hypertargetblue{mr3i}{\mr{3i}} \fi To factory reset the camera, the owner sends the request to the camera, timestamped and authenticated with the secret key $SK_o$ of the owner to verify the legitimacy of the request. Then, both the camera and owner's smartphone delete the key tree $\mathcal{K}$ they have access to, returning both devices to an uninitialized state. See \autoref{Appendix:deletion} for more details.

Note that with this mechanism, delegatees may still know some decryption keys that were deleted from the owner's device. As discussed in our threat model (see \autoref{ThreatModel}), once granted access to a video, a party is not prevented anyway from having already shared or downloaded the video. 

\subsection{Access Recovery}
\label{recovery}

\ifreviews \hypertargetblue{mr3j}{\mr{3j}} \fi In case owners lose access to their smartphone, they must be able to recover access to the system. However, in a privacy-preserving system where no third party is trusted, achieving this is nontrivial. To solve this, we create \textit{escrow material} during the initialization step that contains all of the information needed by the owners to recover access to their system. The escrow material is encrypted and stored on the camera. We note that the escrow material is protected by an unrecoverable passphrase only known by the owner and is therefore not restricted to being stored on the camera. The escrow material gives access to the following secrets :

\begin{itemize}
    \item The \textit{owner's asymmetric key pair} $(SK_o,PK_o)$ encrypted with a randomly generated key of size \SI{128}{bits} which representation in hexadecimal corresponds to the passphrase displayed to the owner during initialization. \ifreviews \hypertargetblue{ra5}{\re[5]{A}} \fi
    \item The \textit{key material} necessary to build the key tree $\mathcal{K}$ (for details about this key tree see \autoref{delegation}) asymmetrically encrypted with the owner's key.
    \item The \textit{asymmetric public key} $PK_c$ of the camera (does not need to be encrypted).
    
\end{itemize}

To recover access, owners use their new smartphone to open a Bluetooth connection to the camera to retrieve the escrow material. Note that no assumption regarding the status of the owner has been made so far; anyone who is in physical range from the camera can request the escrow material. However, only the owners have knowledge of the recovery passphrase and can use it to decrypt the escrow material. Once decrypted, the owner assumes control of the system from the new smartphone and can reconstruct the corresponding key tree $\mathcal{K}$ to retrieve their video footage. More details can be found in \autoref{Appendix:recovery}.

\section{Evaluation}
\label{Evaluation}

The goal of this section is to evaluate the effectiveness and efficiency of \cactus{} with respect to four metrics: security, privacy, usability, and performance. To this end, we evaluate three research questions:

\begin{itemize}

    \item \textbf{RQ1:} Does \cactus{} enforce our privacy requirements, while offering the same feature set found in commercial systems?

    \item \textbf{RQ2:} Is \cactus{} easy to use for end users?
    
    \item \textbf{RQ3:} Can \cactus{} operate at sufficient resolution, frame rate, and latency to meet the needs of smart camera system owners?

\end{itemize}

\noindent\textbf{Experimental Setup.} Experiments were performed using a \textit{Raspberry Pi 4 Model B} with 2GB of RAM, equipped with a camera module. Our implementation of the \cactus{} camera system is written in C, while the paired mobile companion application was written in Java and installed on a \textit{Nokia 4.2} smartphone, running Android 10. Further evaluation setup details are described in \autoref{Appendix:miscellanea}.

\subsection{Privacy and Security Analysis (RQ1)}
\label{privacy_security_analysis}

 \ifreviews \hypertargetblue{mr3k}{\mr{3k}} \fi We begin by analyzing how \cactus{} achieves the security and privacy properties discussed in \autoref{privacy}, as well as related practical concerns.

\noindent\textbf{Confidentiality.} \cactus{} aims to protect user data from being read by unauthorized third parties. As such, owners manage access to the decryption keys, and any administrative actions must be authorized by them. Further, this encryption is performed \textit{end-to-end}: data is encrypted locally at the camera before being stored in the cloud and decrypted locally at the smartphone after being retrieved. As a result, parties who have not been delegated access through key sharing are unable to view the encrypted frames, protecting confidentiality of the footage, which ensures the right to not be seen.

\noindent\textbf{Complete Mediation.} The key-tree construction of \cactus{} ensures that access delegation is performed at a fine-grained level. Because this delegation must be performed by the owner (who holds the seed key of the key tree), no other party is able to grant access to footage, thus providing complete mediation of access. The key tree ensures this mediation cryptographically: a user without the seed or delegated key is unable to decrypt footage regardless of access control on the encrypted data, assuming the cryptographic primitives are secure. This enforces the right to sole ownership.

\noindent\textbf{Data Deletion.} An emerging property of the previous two properties is that the owner has sole ownership and control over the encryption/decryption keys. Thus, if the owner decides they want to delete all stored videos in the cloud, they can simply delete their keys, which makes it impossible for them to view the corresponding encrypted footage stored in the cloud, even in case of coercion. This ensures owners' right to be forgotten.

\begin{figure*}[!ht]
  \centering
  \includegraphics[width=\textwidth]{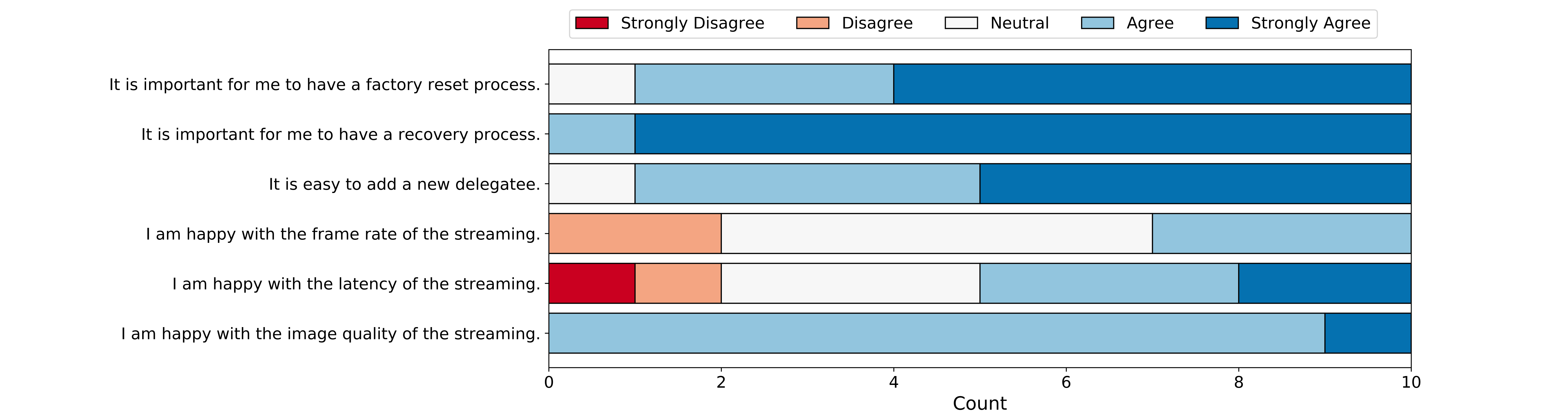}
  \caption{Likert scale evaluation of different statements made by the 10 participants during the functional user evaluation of \cactus{}.}
  \label{fig:likert_all_questions}
\end{figure*}

\subsection{Functional User Evaluation (RQ2)}

We performed a functional user evaluation of our implementation of \cactus{} with ten participants. Studies have shown that this group size is sufficient (twice the minimum number) to identify most of the issues within a system design \cite{experience_why_nodate,nielsen_mathematical_1993}. The goals were to assess the functional ease of use of \cactus{} and identify what aspects could be improved further. To this end, participants were asked to initialize the system and use the different functionalities as if they had just bought it: view live and recorded streams, share videos with a delegatee, recover access after simulating the loss of their smartphone, and reset the system. The detailed protocol that was followed is described in \autoref{Appendix:user_protocol}. All institutional requirements were met for this functional user evaluation of \cactus{}. We obtained approval from the Institutional Review Board (IRB) of our university and a consent form was signed by the participants at the beginning of their session.

Out of the ten participants, four were female and six male, seven of them were between 20-25 years old and the three between 45-55 years old.

\noindent
\ifreviews \hypertargetblue{mr4}{\mr{4}} \fi \textbf{Goals and Limitations.} Our modest goals through this functional user evaluation of \cactus{} were to assess whether guided users can perform the functions of the system that we laid out and identify where our proof-of-concept of the implementation fell short (with respect to performance or feature design) and how it could be improved. As a consequence, our participants knew they were evaluating \cactus{}. Thus, our functional evaluation differs from traditional usability studies that may compare different system designs to identify which is the most usable, has the best interface, or what UX options are optimal for ensuring that users understand a specific privacy concept of the system. We defer a more comprehensive usability study to future work.


\noindent
\ifreviews \hypertargetblue{ra8a}{\re[8a]{A}} \fi \textbf{User Interface.} Overall, the participants thought that the interface of the implemented smart camera system was simple to understand and navigate through. They liked that for every process there were step-by-step instructions displayed to them. They noted that these directions were clear, straightforward, and self-explanatory.

\noindent
\textbf{Secure and Authenticated Pairing.} Despite that pairing requires both setting up a Bluetooth connection and scanning QR codes, the participants found the process simple and easy. They expressed that it was more straightforward than steps they had to perform with other systems. They also felt that using both Bluetooth and a visual channel was more secure even if they did not explicitly always know why.

\noindent
\ifreviews \hypertargetblue{ra6}{\re[6]{A}} \fi\textbf{Quality of the Footage.} As shown in \autoref{fig:likert_all_questions}, the participants agreed that the image quality of the video stream (\SI{480}{p}) and the frame rate (\SI{10}{fps}) were sufficient for security and surveillance purposes. However, some believed that the latency of the system was a bit of a downside as it tended to defeat the initial purpose of being able to monitor what was recorded by the smart camera system in real-time. See \autoref{Discussion} for potential optimizations to \cactus{}.

\noindent
\textbf{Granularity Options.} The participants were impressed by the granularity with which they could view specific segments of video footage (up to the second), but believed that the same granularity option for delegation was not necessary; they stated that the primary use case would be to share access for several hours or days. \ifreviews \hypertargetblue{ra1}{\re[1]{A}} \fi  This concern can be addressed by displaying only up to the minute and adding an option for more precision in the settings of the application for instance. The majority of the participants also expressed desire to have an option to quickly delegate unlimited access, but then figured out that they would not be able to revoke such access without having to reset the system.

\noindent
\textbf{Access Recovery and Factory Reset.} Regarding the recovery process, the participants were divided on whether the recovery passphrase should be randomly generated by the system or if it should give the owner the opportunity to choose it. They felt that they could lose the recovery passphrase or forget about it if they did not choose it. However, they agreed that as this passphrase allows to recover full access to the system, it might be less secure to let the owners pick their own. One participant remarked that the ability to recover access without needing to trust a third party (i.e., without using a recovery email address for instance) was interesting.

As shown on \autoref{fig:likert_all_questions}, most participants found it important to be able to recover access to and factory reset their system, as oftentimes owners may want to recover access (to retrieve videos) before factory resetting it. The participants found the recovery and factory reset processes easy to perform. 

\noindent
\textbf{Missing Features.} Participants expressed that they would like to see the following features implemented: motion detection-triggered recording, remote pairing and delegation, two-way audio support to listen in on and remotely speak through devices, password-locked smartphone application for additional security, and application availability across different platforms (e.g., Android, iOS, and web interface). See \autoref{Discussion} for details about such extensions to \cactus{}.

\subsection{Performance Evaluation (RQ3)}

We now evaluate the streaming performance of \cactus{}, focusing our efforts on three key metrics: latency (delay from time of recording), stream image resolution, and frame rate. As a baseline, commercial systems achieve frame rates of \SI{30}{fps} in \SI{1080}{p} (\SI{1920}{} x \SI{1080}{pixels}), with a latency of several milliseconds. However, we note that these systems have been largely optimized to be sold to consumers, and they do not guarantee similar privacy-preserving features as \cactus{}. We discuss potential optimizations to \cactus{} in \autoref{Discussion}.

\noindent
\textbf{System Latency.} For a video resolution of \resquality{} we obtained a latency of \reslatency{} at a frame rate of \resfps{}.

\noindent
\textbf{System Bottlenecks.} \ifrevision \hypertargetblue{mr2}{\mr{2}}\fi
Next, we measure the effect that each phase of streaming has on the stream latency. The values have been averaged over 1,000 frames. For this evaluation, we picked the same parameters as in \autoref{delegation} for a worst-case scenario (i.e., $d_{\mathcal{K}} = \SI{32}{}$ and $\delta_{\mathcal{K}} = \SI{10}{s}$), to show the performance baseline that can be expected from \cactus{}. \ifreviews \hypertargetblue{ra7}{\re[7]{A}} \fi Table \ref{tab:time_delay_breakdown} shows the results when the camera device is recording for a video quality of \SI{480}{p}. \ifreviews \hypertargetblue{ra4c}{\re[4c]{A}} \fi As shown, the largest contributor to the latency is the upload/download of the encrypted frames to/from the cloud storage server during live stream. As previously stated, we have not optimized this specific point in our implementation of \cactus{} -which would have the same contribution to latency in a system not using encryption techniques- but we discuss means to do so in \autoref{Discussion}. Note that within the same epoch the same key is used, but between epochs, we need to derive the new key, that is why the standard deviation is larger than the average for the key extraction.

\begin{table}[ht]
\centering
\begin{tabular}{llll}
\hline
Device & Operation & Delay (ms) & $\sigma$ (ms) \\ \hline
\multirow{5}{*}{Camera} & Key Extraction & \SI{0.05}{ms} & \SI{0.2}{ms}\\
& Frame Encryption & \SI{2.8}{ms} & \SI{1.0}{ms}\\
& Hash & \SI{1.9}{ms} & \SI{0.7}{ms}\\
& Signature & \SI{8.8}{ms} & \SI{2.9}{ms}\\
& Upload & \SI{510}{ms} & \SI{420}{ms}\\\hline
\multirow{5}{*}{Smartphone} & Download & \SI{420}{ms} & \SI{160}{ms}\\
& Key Extraction & \SI{0.02}{ms} & \SI{0.1}{ms}\\
& Frame Decryption & \SI{0.4}{ms} & \SI{0.5}{ms}\\
& Hash Verification & \SI{1.9}{ms} & \SI{0.4}{ms}\\
& Signature Verification & \SI{0.5}{ms} & \SI{0.5}{ms}\\\hline
\end{tabular}
\caption{Time delay averaged over 1,000 frames and standard deviation $\sigma$ for each live stream operation, while recording with a resolution of \SI{480}{p}.}
\label{tab:time_delay_breakdown}
\end{table}
\section{Discussion}
\label{Discussion}

In the following, we discuss potential optimizations for improving the performance of \cactus{}, practical considerations for deploying it as a commercial system, and extensions of our approach to other devices.

\subsection{Improving Latency}
We identify several components that could be improved to reduce the overall latency of the system, namely; (1) cryptographic accelerators, (2) network relays, (3) streaming libraries, and (4) video compression techniques. Such improvements are described below.

As discussed in \autoref{Evaluation}, our implementation used a Raspberry Pi, wherein the (relatively weak) CPU was responsible for handling all processing, including encryption. Since cryptographic operations dominate many of the features in \cactus{}, we expect substantial gains in latency by leveraging dedicated cryptographic accelerators as seen in many other crypto-dominated applications, such as in IoT~\cite{kietzmann_performance_2021}. 

Streaming live video can be demanding on the network. Specifically, popular video streaming platforms (such as Netflix, Hulu, and Disney Plus) employ a variety of techniques to bring video data as close to the user as possible. These techniques often take the form of caches, content delivery networks, or dedicated network infrastructure designed to serve high-bandwidth content quickly~\cite{florance_about_2016, SALAH202093,adhikari2014measurement}. Naturally, these techniques could substantially improve the performance of \cactus{} to be even closer to commercial-grade systems. 

\ifreviews \hypertargetblue{rb8}{\re[8]{B}}  \fi The smart camera systems available today use optimized streaming protocols to deliver video content quickly~\cite{adhikari2014measurement}. Given that it was necessary for us to implement our video streaming protocol from scratch (to support our encryption and key rotation schemes), we could see further improvements by augmenting current protocols to support our design. In a similar vein, our current implementation operates at the frame-level, while commercial systems operate at the block-level (and thus exploit compression algorithms commonly used in video streaming applications~\cite{apostolopoulos2002video}). Moreover, popular techniques such as adaptive video playback or frame dropping could also be used to improve the throughput of \cactus{}. We defer such improvements to future work.

\begin{table*}[!t]
\centering
\begin{tabular}{ccccc}
\hline
& \multicolumn{2}{c}{\textit{Right to not be seen}} & \textit{Right of sole ownership} & \textit{Right to be forgotten}  \\
Camera System & Video Encryption & Owner Controls Access & Only the Owner is Trusted & Video Deletion   \\ \hline
PrivacyCam~\cite{chattopadhyay_privacycam_2007} & \redcheck~(ROI only) & \redcheck & \redcheck & \redcheck  \\
TrustCAM~\cite{winkler_trustcam_2010} & \redcheck~(ROI only) & \redcheck & \redcheck~(trusts a central station) & \redcheck \\
TrustEYE.M4~\cite{winkler_secure_2015} & \greencheck & \redcheck & \greencheck & \redcheck \\
SoC-based~\cite{haider_private_2017}   & \greencheck & \redcheck  & \redcheck~(relies on a Trusted Authority) & \redcheck \\
Signcryption~\cite{ullah_smart_2017}  & \greencheck  & \redcheck & \redcheck~(uses a Key Distribution Center) & \redcheck \\
Pinto~\cite{yu_pinto_2018} & \redcheck~(ROI only) & \redcheck & \redcheck & \redcheck  \\
\cactus{} & \greencheck & \greencheck & \greencheck & \greencheck\\ \hline
\end{tabular}
\caption{Comparison of privacy guarantees of smart camera systems proposed in the literature (ROI stands for Region Of Interest).}
\label{tab:features_comparison}
\end{table*}

\subsection{Deploying CaCTUS as a Commercial System}
\label{commercialsystem}
\ifreviews \hypertargetblue{ra8b}{\re[8b]{A}} \fi Here, we highlight some challenges (motivated by notable features in commercial systems and suggestions from our functional user evaluation of \cactus{}) that should be addressed to realize commercial implementations of \cactus{} without compromising any of our privacy goals. 

\noindent
\textbf{Relaxing Proximity.} To uphold our privacy goals, delegatees need to be within local proximity of the owners (e.g., to perform delegation and pairing through the QR codes). While this was not an area of concern in the functional user evaluation of \cactus{} since participants were always close to the camera system, this can be challenging if users wish to delegate access remotely. We did conceptually derive a scheme to support this capability, while upholding our privacy goals. Specifically, \cactus{} could be extended to asymmetrically\footnote{Clearly, how the public and private keys are computed and transferred between the owner and the remote delegatee needs to be done in a security-preserving manner.} encrypt the keys needed by delegatees and upload them to the cloud storage or use another third party service like email, from where the remote delegatee could retrieve, decrypt, and then use these keys to access the video stream for some predetermined period of time. 

\noindent
\ifreviews \hypertargetblue{rb9}{\re[9]{B}} \fi
\textbf{Revoking Access to Unrecorded Videos.} Note that a similar scheme to the one described in the previous paragraph could also be used to allow owners to revoke access to delegatees to unrecorded videos without needing to factory reset the system. The sharing of the keys only needs to be done periodically at a certain interval that could be configured by the owner for each delegatee.

\noindent
\textbf{Supporting Audio.} In our current implementation, we focused on supporting video data only, and not audio. However, the system design does not prohibit extensions to support audio-video recording and streaming. The extension is straightforward: we can treat some audio sample as a ``frame'', as it is done for video, and apply similar encryption operations. For sampling, a Linear Predictive Coding (LPC), which is widely deployed by telephone companies for speech encoding and processing, could be used~\cite{scorletti:cel-00673929}. Moreover, LPC can even be leveraged to transmit audio in reverse; that is, be applied so the users could speak into their smartphones and have the audio replayed through the camera, as commonly seen in commercial systems. 

\noindent
\ifreviews \hypertargetblue{rb10a}{\re[10a]{B}} \fi \ifrevision \hypertargetblue{mr4b}{\mr{4b}}  \hypertargetblue{mr3f}{\mr{3f}}\fi \textbf{Supporting Motion Detection.} Many commercial systems support forms of motion detection to notify users of events they may be interested in viewing. We find this feature valuable for both users as well as the camera system in that video recording need only be saved if motion was detected\footnote{To mitigate side-channel attacks and to not leak behavior patterns, meaningless data could be randomly uploaded to the cloud.}. \cactus{} could support this capability through addition of a physical hardware sensor or a software solution. This addition would save significant space in cloud storage systems and reduce the cost related to storage. In a similar vein older data could be overwritten at a fixed interval that could be configured by the owner. 

\noindent
\textbf{Physical Security.} Finally, as explained in our threat model, we did not take into consideration physical tampering attacks against the camera device. Techniques such as package design so that the device is tamper-proof should be explored. Specifically, local storage, battery, reset buttons, serial ports, firmware, etc., should be protected to prevent physical access by untrusted parties.

\subsection{Extending to Other Devices}
\ifreviews \hypertargetblue{rb10b}{\re[10b]{B}} \fi
The protocols used in \cactus{} could be easily extended to other devices recording time-stamped data streams such as; temperature, humidity level, heartbeat monitoring, audio, power usage, etc. As video stream is the most bandwidth-intensive of these applications, performance for such devices is expected to be even better. Note that in the initialization step of \cactus{}, we leverage the fact that both devices (the camera and the smartphone) have a camera sensors, however, this may not be the case for other IoT devices, we defer to the \textit{Seeing-Is-Believing} (SiB) technique introduced by McCune, Perrig, and Reiter~\cite{mccune_seeing-is-believing_2005} for a discussion of the guarantees provided in such a case.

\section{Related Work}
\label{Related Work}

While this work is the first to examine a smart camera system that is privacy-preserving, there is already a large body of research studying the security of smart camera systems. Here, we detail the gaps in prior work and how \cactus{} addresses them. Refer to \autoref{tab:features_comparison} for a comparison of the privacy guarantees of these systems.

Alharbi and Aspinall introduced a security analysis framework for IoT smart cameras that weighs the threat of significant risks (e.g., unencrypted video streaming) across various platforms~\cite{alharbi_iot_2018}. However, it focuses on the security of the camera device and only partially addresses some vulnerabilities of other components of the system. For instance, the authors do not discuss the use of cloud storage to remotely access recordings, nor procedures for securely pairing a smartphone and camera. \cactus{} provides an end-to-end secure solution starting from camera initialization to protect the secrecy and integrity of both communications and data.

Haider and Rinner proposed a SoC-based smart camera that uses physically unclonable functions (PUFs) to generate encryption keys~\cite{haider_private_2017}. The keys are used to encrypt video frames at the camera before storing them in remote cloud storage, thus removing the requirement that the owner trust the cloud service provider. However, this approach has several limitations. First, a trusted authority is required to create camera device fingerprints during key generation; in a commercial system, the trusted authority will likely be the manufacturer, whom in general is not trusted by the camera owner. Moreover, the key extraction procedure using PUFs only produces a fixed number of encryption keys, which does not align with the feature-set typically desired by smart-camera owners---e.g., being able to delegate camera access to other people with fine granularity. Finally, as the fingerprint is physically embedded into the hardware, camera owners will likely need to get a brand new device if the encryption keys are leaked. \cactus{} addresses these shortcomings by making the owner (i.e., their smartphone) the root of trust: they store the secrets and share expendable keys with both the camera and delegatees. This simultaneously gives control back to the owner while enabling delegation and system reset without specialized (or new) hardware.

Winkler and Rinner introduced TrustEYE.MP4~\cite{winkler_secure_2015}, a monitoring framework that provides similar secrecy and integrity guarantees envisioned by \cactus{}. However, as the same key is used to encrypt all the videos, they do not address the unique challenges in delegating access (e.g., how to share and revoke access to video data at particular timescales) or deleting videos. Similarly, Ullah, Rinner, and Marcenaro proposed using signencryption~\cite{ullah_smart_2017} to encrypt and sign video frames at once, but they also did not consider the challenges related to video deletion and delegation. \cactus{} addresses these issues by storing the system secrets at the owner's smartphone and using a secure key rotation scheme to enable fine-grained delegation.

Finally, PrivacyCam~\cite{chattopadhyay_privacycam_2007}, TrustCAM~\cite{winkler_trustcam_2010}, and Pinto~\cite{yu_pinto_2018} do not fully address the privacy challenges in a smart camera system \ifrevision \hypertargetblue{mr1a}{\mr{1a}} \fi as they attempt to solve a problem of a different nature and setting. These systems try to protect the anonymity of people or of vehicle license plates recorded in public spaces by detecting privacy sensitive regions and selectively encrypting or blurring them while leaving the rest of the video frame unperturbed. However, even this approach is limited in protecting anonymity: people can still be identified through their clothes or actions, and blurred videos still disclose behavior patterns such as when users are at home. Moreover, the system does not address data privacy of the video recordings. 
\section{Conclusion}
\label{Conclusion}

\ifrevision \hypertargetblue{mr1c}{\mr{1c}} \fi
This paper presented \cactus{}, a smart camera system with popular commercial features that \textit{returns control of videos to the users}. In \cactus{}, we provide the owners with full control over their system, isolation and protection of the access to the video footage, deletion and factory reset, as well as peer-to-peer and fine-grained delegation. We leverage physical and direct pairing for system initialization (that is, without relying on trusting third parties), performance-aware cryptographic algorithms to support video streaming, key rotation and management through a binary key tree to provide video deletion, factory reset, and fine-grained (i.e., on the order of seconds) peer-to-peer delegation of video footage. Additionally, \cactus{} satisfies performance requirements necessary to execute the most popular commercial functionalities of smart camera systems, such as live streaming with \reslatency{} of latency at a frame rate of \resfps{} and resolution of \resquality{}, while protecting the users' privacy. \cactus{} serves as an existence proof that smart camera systems need not compromise the privacy of users to be afforded the modern capabilities that commercial systems offer today.
\section{Acknowledgment}
\iffinal
We would like to thank Christie Warren for her help in the design of the user interface of the smartphone application for this project as well as Dr. Hanrahan for his valuable feedback on the protocol of the functional user evaluation of \cactus{}. We also sincerely thank all the users that have participated in the test and evaluation sessions of our implementation of CaCTUs. 

\textbf{Funding acknowledgment:} This material is based upon work supported by, or in part by, the National Science Foundation under Grant No. CNS-1805310 and Grant No. CNS-1564105, and the U.S. Army Research Laboratory and the U.S. Army Research Office under Grant No. W911NF-19-1-0374. Any opinions, findings, and conclusions or recommendations expressed in this publication are those of the author(s) and do not necessarily reflect the views of the National Science Foundation, or the U.S. Government. The U.S. Government is authorized to reproduce and distribute reprints for government purposes notwithstanding any copyright notation hereon.

\else 
\texttt{ANONYMIZED FOR SUBMISSION}
\fi

\ifrevision \hypertargetblue{mr1d}{\mr{1d}} \fi

\bibliographystyle{pets_style}
\bibliography{biblio}

\begin{thebibliography}{10}
\providecommand{\url}[1]{\texttt{#1}}
\providecommand{\urlprefix}{URL: }
\expandafter\ifx\csname urlstyle\endcsname\relax
  \providecommand{\doi}[1]{doi:\discretionary{}{}{}#1}\else
  \providecommand{\doi}{doi:\discretionary{}{}{}\begingroup
  \urlstyle{rm}\Url}\fi

\bibitem{adhikari2014measurement}
V.~K. Adhikari et~al.
\newblock Measurement study of Netflix, Hulu, and a tale of three CDNs.
\newblock IEEE/ACM Transactions on Networking, 23 (2014)(6):1984--1997.

\bibitem{alharbi_iot_2018}
R.~Alharbi and D.~Aspinall.
\newblock An {IoT} analysis framework: {An} investigation of {IoT} smart
  cameras' vulnerabilities.
\newblock In: Living in the {Internet} of {Things}: {Cybersecurity} of the
  {IoT} - 2018, pages 1--10 (2018).
\newblock \doi{10.1049/cp.2018.0047}.

\bibitem{apostolopoulos2002video}
J.~G. Apostolopoulos, W.-t. Tan, and S.~J. Wee.
\newblock Video streaming: Concepts, algorithms, and systems.
\newblock HP Laboratories, report HPL-2002-260,  (2002).

\bibitem{arlo_arlo_2021}
Arlo.
\newblock Arlo - {Investor} {Relations} (2021).
\newblock \urlprefix\url{https://investor.arlo.com/ir-home/default.aspx}.
\newblock Last Accessed: 2021-04-06.

\bibitem{biddle_for_2019}
S.~Biddle.
\newblock For {Owners} of {Amazon}’s {Ring} {Security} {Cameras}, {Strangers}
  {May} {Have} {Been} {Watching} {Too}.
\newblock The Intercept,  (2019).
\newblock
  \urlprefix\url{https://theintercept.com/2019/01/10/amazon-ring-security-camera/}.
\newblock Last Accessed: 2020-09-08.

\bibitem{european_data_protection_board_guidelines_2019}
E.~D.~P. Board.
\newblock Guidelines 3/2019 on processing of personal data through video
  devices (2019).
\newblock
  \urlprefix\url{https://edpb.europa.eu/sites/edpb/files/consultation/edpb_guidelines_201903_videosurveillance.pdf}.
\newblock Last Accessed: 2021-04-19.

\bibitem{brewster_smart_2018}
T.~Brewster.
\newblock Smart {Home} {Surveillance}: {Governments} {Tell} {Google}'s {Nest}
  {To} {Hand} {Over} {Data} 300 {Times}.
\newblock Forbes,  (2018).
\newblock
  \urlprefix\url{https://www.forbes.com/sites/thomasbrewster/2018/10/13/smart-home-surveillance-governments-tell-googles-nest-to-hand-over-data-300-times/}.
\newblock Last Accessed: 2020-09-08.

\bibitem{bridges_amazons_2021}
L.~Bridges.
\newblock Amazon’s {Ring} is the largest civilian surveillance network the
  {US} has ever seen {\textbar} {Lauren} {Bridges}.
\newblock The Guardian,  (2021).
\newblock
  \urlprefix\url{http://www.theguardian.com/commentisfree/2021/may/18/amazon-ring-largest-civilian-surveillance-network-us}.
\newblock Last Accessed: 2021-05-18.

\bibitem{cameron_amazon_2019}
D.~Cameron.
\newblock Amazon {Is} {Marketing} {Face} {Recognition} to {Police}
  {Departments} {Partnered} {With} {Ring}: {Report}.
\newblock Gizmodo,  (2019).
\newblock
  \urlprefix\url{https://gizmodo.com/amazon-is-marketing-face-recognition-to-police-departme-1839073749}.
\newblock Last Accessed: 2020-09-08.

\bibitem{cameron_rings_2019}
D.~Cameron and D.~Mehrotra.
\newblock Ring’s {Hidden} {Data} {Let} {Us} {Map} {Amazon}'s {Sprawling}
  {Home} {Surveillance} {Network}.
\newblock Gizmodo,  (2019).
\newblock
  \urlprefix\url{https://gizmodo.com/ring-s-hidden-data-let-us-map-amazons-sprawling-home-su-1840312279}.
\newblock Last Accessed: 2020-01-13.

\bibitem{chattopadhyay_privacycam_2007}
A.~Chattopadhyay and T.~E. Boult.
\newblock {PrivacyCam}: a {Privacy} {Preserving} {Camera} {Using} {uCLinux} on
  the {Blackfin} {DSP}.
\newblock In: 2007 {IEEE} {Conference} on {Computer} {Vision} and {Pattern}
  {Recognition}, pages 1--8 (2007).
\newblock \doi{10.1109/CVPR.2007.383413}.

\bibitem{cimpanu_hackers_2019}
C.~Cimpanu.
\newblock Hackers keep dumping {Ring} credentials online 'for the giggles'.
\newblock ZDNet,  (2019).
\newblock
  \urlprefix\url{https://www.zdnet.com/article/hackers-keep-dumping-ring-credentials-online-for-the-giggles/}.
\newblock Last Accessed: 2020-01-13.

\bibitem{deahl_ring_2019}
D.~Deahl.
\newblock Ring let employees watch customer videos, claim reports.
\newblock The Verge,  (2019).
\newblock
  \urlprefix\url{https://www.theverge.com/2019/1/10/18177305/ring-employees-unencrypted-customer-video-amazon}.
\newblock Last Accessed: 2020-09-08.

\bibitem{diffie_new_1976}
W.~Diffie and M.~Hellman.
\newblock New directions in cryptography.
\newblock IEEE Transactions on Information Theory, 22 (1976)(6):644--654.
\newblock \doi{10.1109/TIT.1976.1055638}.

\bibitem{dirks_video_2009}
B.~Dirks et~al.
\newblock Video for {Linux} {Two} {API} {Specification} (2009).
\newblock
  \urlprefix\url{https://www.linuxtv.org/downloads/legacy/video4linux/API/V4L2_API/spec-single/v4l2.html}.
\newblock Last Accessed: 2021-03-21.

\bibitem{florance_about_2016}
K.~Florance.
\newblock About {Netflix} - {How} {Netflix} {Works} {With} {ISPs} {Around} the
  {Globe} to {Deliver} a {Great} {Viewing} {Experience}.
\newblock About Netflix,  (2016).
\newblock
  \urlprefix\url{https://about.netflix.com/en/news/how-netflix-works-with-isps-around-the-globe-to-deliver-a-great-viewing-experience}.
\newblock Last Accessed: 2021-05-28.

\bibitem{flores_bad_2020}
Y.~Flores.
\newblock Bad {Neighbors}? {How} {Amazon}’s {Ring} {Video} {Surveillance}
  {Could} be {Undermining} {Fourth} {Amendment} {Protections} (2020).
\newblock
  \urlprefix\url{https://www.californialawreview.org/amazon-ring-undermining-fourth-amendment/}.
\newblock Last Accessed: 2021-05-18.

\bibitem{gennaro_how_2001}
R.~Gennaro and P.~Rohatgi.
\newblock How to {Sign} {Digital} {Streams}.
\newblock Information and Computation, 165 (2001)(1):100--116.
\newblock \doi{10.1006/inco.2000.2916}.

\bibitem{greenberg_two_2021}
A.~Greenberg.
\newblock Two {Cases}' {Lessons}: {If} {Cops} {Don}'t {Know} {What} {You}
  {Encrypted}, {They} {Can}'t {Make} {You} {Decrypt} {It} (2021).
\newblock
  \urlprefix\url{https://www.forbes.com/sites/andygreenberg/2012/02/24/two-cases-lessons-if-cops-dont-know-what-you-encrypted-they-cant-make-you-decrypt-it/}.
\newblock Last Accessed: 2021-08-02.

\bibitem{guariglia_lapd_2021}
M.~Guariglia and M.~Maas.
\newblock {LAPD} {Requested} {Ring} {Footage} of {Black} {Lives} {Matter}
  {Protests}.
\newblock Electronic Frontier Foundation,  (2021).
\newblock
  \urlprefix\url{https://www.eff.org/deeplinks/2021/02/lapd-requested-ring-footage-black-lives-matter-protests}.
\newblock Last Accessed: 2021-05-11.

\bibitem{haider_private_2017}
I.~Haider and B.~Rinner.
\newblock Private {Space} {Monitoring} with {SoC}-{Based} {Smart} {Cameras}.
\newblock In: 2017 {IEEE} 14th {International} {Conference} on {Mobile} {Ad}
  {Hoc} and {Sensor} {Systems} ({MASS}), pages 19--27 (2017).
\newblock \doi{10.1109/MASS.2017.15}.

\bibitem{herrman_whos_2020}
J.~Herrman.
\newblock Who’s {Watching} {Your} {Porch}?
\newblock The New York Times,  (2020).
\newblock
  \urlprefix\url{https://www.nytimes.com/2020/01/19/style/ring-video-doorbell-home-security.html}.
\newblock Last Accessed: 2021-04-06.

\bibitem{huseman_huseman_2020}
B.~Huseman.
\newblock Huseman reply to {Wyden}, {Markey}, {Van} {Hollen}, {Coons}, {Peters}
  letter about {Ring}’s {Data} {Security} {Practices} (2020).
\newblock
  \urlprefix\url{https://regmedia.co.uk/2020/01/08/ringsenateresponse.pdf}.
\newblock Last Accessed: 2021-04-06.

\bibitem{katz_introduction_2014}
J.~Katz and Y.~Lindell.
\newblock Introduction to {Modern} {Cryptography}, {Second} {Edition} (2014).
\newblock Chapman \& Hall/CRC.

\bibitem{keck_amazons_2019}
C.~Keck.
\newblock Amazon's {Ring} {Security} {Cameras} {May} {Have} {Let} {Employees}
  {Spy} on {Customers}: {Report}.
\newblock Gizmodo,  (2019).
\newblock
  \urlprefix\url{https://gizmodo.com/amazons-ring-security-cameras-may-have-let-employees-sp-1831658669}.
\newblock Last Accessed: 2020-09-08.

\bibitem{kietzmann_performance_2021}
P.~Kietzmann, L.~Boeckmann, L.~Lanzieri, T.~C. Schmidt, and M.~W\"{a}hlisch.
\newblock A {Performance} {Study} of {Crypto}-{Hardware} in the {Low}-end
  {IoT}.
\newblock In: Proceedings of the 2021 {International} {Conference} on
  {Embedded} {Wireless} {Systems} and {Networks}, pages 79--90 (2021).
\newblock \doi{10.5555/3451271.3451279}.

\bibitem{kocher_complexity_2011}
P.~Kocher.
\newblock Complexity and the challenges of securing {SoCs}.
\newblock In: 2011 48th {ACM}/{EDAC}/{IEEE} {Design} {Automation} {Conference}
  ({DAC}), pages 328--331 (2011).

\bibitem{kraus_ring_2019}
R.~Kraus.
\newblock Ring watched your kids trick or treat and then bragged about it.
\newblock Mashable,  (2019).
\newblock
  \urlprefix\url{https://mashable.com/article/ring-halloween-surveillance/}.
\newblock Last Accessed: 2021-05-11.

\bibitem{kravets_indefinite_2016}
D.~Kravets.
\newblock Indefinite prison for suspect who won’t decrypt hard drives, feds
  say.
\newblock Ars Technica,  (2016).
\newblock
  \urlprefix\url{https://arstechnica.com/tech-policy/2016/05/feds-say-suspect-should-rot-in-prison-for-refusing-to-decrypt-drives/}.
\newblock Last Accessed: 2021-05-31.

\bibitem{lecher_ring_2019}
C.~Lecher.
\newblock Ring reportedly outed camera owners to police with a heat map.
\newblock The Verge,  (2019).
\newblock
  \urlprefix\url{https://www.theverge.com/2019/12/3/20993814/ring-user-location-heat-map-police-privacy-tool-camera-owners}.
\newblock Last Accessed: 2020-09-08.

\bibitem{california_state_legislature_title_2018}
C.~S. Legislature.
\newblock {TITLE} 1.81.5. {California} {Consumer} {Privacy} {Act} of 2018
  [1798.100 - 1798.199.100] (2018).

\bibitem{california_state_legislature_california_2020}
C.~S. Legislature.
\newblock The {California} {Privacy} {Rights} {Act} of 2020 (2020).

\bibitem{mccune_seeing-is-believing_2005}
J.~McCune, A.~Perrig, and M.~Reiter.
\newblock Seeing-is-believing: using camera phones for human-verifiable
  authentication.
\newblock In: 2005 {IEEE} {Symposium} on {Security} and {Privacy} ({S} {P}'05),
  pages 110--124 (2005).
\newblock \doi{10.1109/SP.2005.19}.

\bibitem{ng_amazons_2019}
A.~Ng.
\newblock Amazon's {Ring} wanted to use 911 calls to activate its video
  doorbells.
\newblock CNET,  (2019).
\newblock
  \urlprefix\url{https://www.cnet.com/home/smart-home/amazons-ring-wanted-to-use-911-calls-to-activate-its-video-doorbells/}.
\newblock Last Accessed: 2021-05-18.

\bibitem{ng_ring_2019}
A.~Ng.
\newblock Ring let police view map of video doorbell installations for over a
  year.
\newblock CNET,  (2019).
\newblock
  \urlprefix\url{https://www.cnet.com/news/ring-gave-police-a-street-level-view-of-where-video-doorbells-were-for-over-a-year/}.
\newblock Last Accessed: 2020-01-13.

\bibitem{experience_why_nodate}
J.~Nielsen.
\newblock Why {You} {Only} {Need} to {Test} with 5 {Users}.
\newblock
  \urlprefix\url{https://www.nngroup.com/articles/why-you-only-need-to-test-with-5-users/}.
\newblock Last Accessed: 2021-08-30.

\bibitem{nielsen_mathematical_1993}
J.~Nielsen and T.~K. Landauer.
\newblock A mathematical model of the finding of usability problems (1993).
\newblock \doi{10.1145/169059.169166}.

\bibitem{house_of_commons_of_canada_bill_2020}
H.~of~Commons~of Canada.
\newblock Bill {C}-11 ({First} {Reading}) (2020).

\bibitem{european_parliament_eprivacy_2009}
E.~Parliament and C.~of~the European~Union.
\newblock {ePrivacy} {Directive} - {Directive} 2009/136/{EC} (2009).

\bibitem{european_parliament_regulation_2016}
E.~Parliament and C.~of~the European~Union.
\newblock Regulation ({EU}) 2016/679 of the {European} {Parliament} and of the
  {Council} of 27 {April} 2016 on the protection of natural persons with regard
  to the processing of personal data and on the free movement of such data, and
  repealing {Directive} 95/46/{EC} ({General} {Data} {Protection} {Regulation})
  ({Text} with {EEA} relevance) (2016).

\bibitem{paul_amazons_2019}
K.~Paul.
\newblock Amazon's doorbell camera {Ring} is working with police – and
  controlling what they say.
\newblock The Guardian,  (2019).
\newblock
  \urlprefix\url{https://www.theguardian.com/technology/2019/aug/29/ring-amazon-police-partnership-social-media-neighbor}.
\newblock Last Accessed: 2020-01-13.

\bibitem{the_constitution_project_guidelines_2007}
T.~C. Project.
\newblock Guidelines for public video surveillance - {A} guide to protecting
  communities and preserving civil liberties (2007).
\newblock
  \urlprefix\url{https://archive.constitutionproject.org/pdf/Video_Surveillance_Guidelines_Report_w_Model_Legislation4.pdf}.
\newblock Last Accessed: 2021-04-19.

\bibitem{ring_ring_2019}
Ring.
\newblock Ring {Video} {Doorbells} {Get} 15+ {Million} {Dings} {This}
  {Halloween} and {Capture} {Cute} {Costumes} and {Fun} {Pranks}.
\newblock The Ring Blog,  (2019).
\newblock
  \urlprefix\url{https://blog.ring.com/neighborhood-stories/ring-video-doorbells-get-15-million-dings-this-halloween-and-capture-cute-costumes-and-fun-pranks/}.
\newblock Last Accessed: 2021-05-18.

\bibitem{ring_active_2021}
Ring.
\newblock Active {Agency} {Map} (2021).
\newblock
  \urlprefix\url{https://www.google.com/maps/d/viewer?mid=1eYVDPh5itXq5acDT9b0BVeQwmESBa4cB}.
\newblock Last Accessed: 2021-05-18.

\bibitem{rivest_cryptographic_1983}
R.~L. Rivest, A.~Shamir, and L.~M. Adleman.
\newblock Cryptographic communications system and method (1983).
\newblock \urlprefix\url{https://patents.google.com/patent/US4405829/en}.
\newblock Last Accessed: 2021-08-18.

\bibitem{ropek_home_2021}
L.~Ropek.
\newblock A {Home} {Security} {Worker} {Hacked} {Into} {Surveillance} {Systems}
  to {Watch} {People} {Have} {Sex}.
\newblock Gizmodo,  (2021).
\newblock
  \urlprefix\url{https://gizmodo.com/a-home-security-worker-hacked-into-surveillance-systems-1846111569}.
\newblock Last Accessed: 2021-01-23.

\bibitem{SALAH202093}
H.~Salah, S.~Zimmermann, and J.~A. {Cabrera G.}
\newblock Chapter 5 - Content distribution (2020).
\newblock \doi{https://doi.org/10.1016/B978-0-12-820488-7.00016-5}.

\bibitem{scorletti:cel-00673929}
G.~Scorletti.
\newblock {Traitement du Signal} (2016).
\newblock \urlprefix\url{https://cel.archives-ouvertes.fr/cel-00673929}.
\newblock Last Accessed: 2021-05-11 (Lecture material in French).

\bibitem{ullah_smart_2017}
S.~Ullah, B.~Rinner, and L.~Marcenaro.
\newblock Smart cameras with onboard signcryption for securing {IoT}
  applications.
\newblock In: 2017 {Global} {Internet} of {Things} {Summit} ({GIoTS}), pages
  1--6 (2017).
\newblock \doi{10.1109/GIOTS.2017.8016279}.

\bibitem{winkler_trustcam_2010}
T.~Winkler and B.~Rinner.
\newblock {TrustCAM}: {Security} and {Privacy}-{Protection} for an {Embedded}
  {Smart} {Camera} {Based} on {Trusted} {Computing}.
\newblock In: 2010 7th {IEEE} {International} {Conference} on {Advanced}
  {Video} and {Signal} {Based} {Surveillance}, pages 593--600 (2010).
\newblock \doi{10.1109/AVSS.2010.38}.

\bibitem{winkler_secure_2015}
T.~Winkler and B.~Rinner.
\newblock Secure embedded visual sensing in end-user applications with
  TrustEYE.M4.
\newblock In: 2015 IEEE Tenth International Conference on Intelligent Sensors,
  Sensor Networks and Information Processing (ISSNIP), pages 1--6 (2015).
\newblock \doi{10.1109/ISSNIP.2015.7106934}.

\bibitem{wyden_wyden_2019}
R.~Wyden, C.~Van~Hollen, E.~Markey, C.~Coons, and G.~Peters.
\newblock Wyden, {Markey}, {Van} {Hollen}, {Coons}, {Peters} {Question}
  {Ring}’s {Data} {Security} {Practices} (2019).
\newblock
  \urlprefix\url{https://www.wyden.senate.gov/news/press-releases/wyden-markey-van-hollen-coons-peters-question-rings-data-security-practices}.
\newblock Last Accessed: 2021-04-06.

\bibitem{wyze_wyze_2018}
Wyze.
\newblock Wyze {Cam} - {Our} {Story} (2018).
\newblock \urlprefix\url{https://wyze.com/our-story}.
\newblock Last Accessed: 2021-04-06.

\bibitem{yu_pinto_2018}
H.~Yu, J.~Lim, K.~Kim, and S.-B. Lee.
\newblock {Pinto}: {Enabling} {Video} {Privacy} for {Commodity} {IoT}
  {Cameras}.
\newblock In: Proceedings of the 2018 {ACM} {SIGSAC} {Conference} on {Computer}
  and {Communications} {Security}, pages 1089--1101 (2018).
\newblock \doi{10.1145/3243734.3243830}.

\end{thebibliography}
\newpage
\appendix
\section{Notation}
\label{Appendix:notation}

\begin{table}[!ht]
\centering
\begin{tabular}[t]{m{2.5cm}m{4.5cm}}
\hline
Symbol & Name \\ \hline
 $||$ & Concatenate function\\
 $\oplus$ & XOR function\\
 $AES256Dec$ & Symmetrical decryption function\\
 $AES256Enc$ & Symmetrical encryption function\\
 $C_i$ & $i$th cipher frame\\
 $check$ & Function to check that corresponding party knows the secret key linked to the public key.\\
 $d_\mathcal{K}$ & Depth key tree \\
 $\delta_\mathcal{K}$ & Epoch size\\
 \textit{escrow material} & Escrow material encrypted with a passphrase\\
 $Extract$ & Key extraction function from $\mathcal{K}$\\
 $F_i$ & $i$th frame \\
 $gen$ & Generate function\\
 $h_i$ & $i$th hash \\
 $h_{PK_d}$ & Hash of the public key of a delegatee\\ 
 $h_{PK_f}$ & Hash of the factory-generated public key of the camera\\ 
 $h_{PK_o}$ & Hash of the  public key of the owner\\ 
 $hash$ & Hash function\\
 $HKDF$ & Hash-based key derivation function\\
 $HMAC$ & Hash-based message authentication code function\\
 $init$ & Initialization function\\
 $IV_{i}$ & $i$th initialization vector \\
 $\mathcal{K}$ & Binary key tree \\
 $k_{i}$ & Symmetric encryption key for $F_i$ \\
 \textit{passphrase} & Passphrase encrypting the escrow material\\
  $RandBytes$ & Random bytes generation function\\
  $RSA$ & Rivest–Shamir–Adleman encryption and signature scheme\\
 \textit{secrets} & Secrets sent by the owner to the camera during the initialization\\
 \textit{seed key} & Encryption material owned by the root node of $\mathcal{K}$\\
 $\sigma$ & Asymmetrical signature of a block of frames\\
 $Sign$ & Asymmetrical signature function\\
 $(SK_{c}, PK_{c})$ & Asymmetric key pair of the camera\\
 $(SK_{d}, PK_{d})$ & Asymmetric key pair of a delegatee\\
 $(SK_{f}, PK_{f})$ & Factory-generated asymmetric key pair of the camera\\
 $(SK_{o}, PK_{o})$ & Asymmetric key pair of the owner\\
  $t_i$ & Timestamp of $i$th frame \\
 $[t_j, t_{j+1})$ & Time interval of the $j$th epoch\\
 $Verify$ & Signature verification function\\
 \textit{wifi credentials} & Wifi credentials of the owner's network\\
 \hline
\end{tabular}%
\caption{Notation used by \cactus{} cryptographic constructions.}
\label{tab:notation}
\end{table}

\section{Experimental Setup}
\label{Appendix:miscellanea}

\noindent
\textbf{Camera Device.} On a \textit{Raspberry Pi 4 Model B Rev 1.1} (Broadcom BCM2711, 1.5 GHz quad-core Cortex-A72 ARM v7 64-bit, 2GB RAM), we used the Video4Linux2 driver \cite{dirks_video_2009} to interface with the camera sensor and capture frames that are then encrypted using OpenSSL3.0\footnote{\url{https://www.openssl.org/}}. The Raspberry Pi Camera Module v2 that we used has a still resolution of 8 Megapixels, a sensor resolution of $3280 \times 2464$ pixels, and supports the three following video modes \SI{1080}{p}/\SI{30}{fps}, \SI{720}{p}/\SI{60}{fps}, and \SI{480}{p}/\SI{90}{fps} (respectively video quality and maximum frame rate).

\noindent
\textbf{Android Smartphone.} We used a \textit{Nokia 4.2} smartphone with Android 10 on which we have installed the implemented application. In this application, we use C native libraries that we have cross-compiled, and C code to download and decrypt the frames. We leveraged the MediaCodec class\footnote{\url{https://developer.android.com/reference/android/media/MediaCodec}} to perform the encoding and decoding of video files, as well as the Quirc\footnote{\url{https://github.com/dlbeer/quirc}} and Bluetooth libraries to perform the pairing.

\noindent
\textbf{Cloud Storage.} An \textit{AWS EC2 t3.small} instance was used to deploy a Nginx web server. Upon request, we serve the list of encrypted frames that were recorded during the time frame specified in the request.

\section{Signing Every Frame}
\label{Appendix:signing}

We can individually sign each transmitted frame with an adaptation of the one-time signatures (also known as hashed signatures) process described by Gennaro and Rohatgi~\cite{gennaro_how_2001}. This approach is more suitable for live streaming, as the receiving device does not need to verify the entire block of $N$ frames before playing a frame. The idea is to incorporate a one-time public key into each frame payload, which is then used to sign the next frame. During initialization, only the first frame is signed asymmetrically, and subsequent frames are faster to sign and verify. Note, however, that there is a trade-off to this approach, as generation of the one-time keys still incurs a computational cost (though this could be reduced by pre-generating keys). We concretely describe this technique, which is implemented using a hash-based key tree:

\begin{enumerate}
    \item Each frame $F_i$ is recorded at timestamp $t_i$.
    \item The corresponding symmetric key $k_{i}$ is extracted from the key rotation scheme $\mathcal{K}$, an initialization vector $IV_{i}$ is randomly generated, and a pair of one-time keys $(PK_{i+1},SK_{i+1})$ is derived.
    \[
    k_{i} = Extract(\mathcal{K}, i)
    \]
    \[
    IV_{i} = RandBytes(16)
    \]
    \[
    (PK_{i+1},SK_{i+1}) = OneTimePair()
    \]
    \item Each frame is then symmetrically encrypted (\textbf{confidentiality}) into the corresponding cipher $C_i$ using the AES algorithm with a 256-bit key in Galois/Counter Mode (GCM) mode. $C_i$ is then concatenated with $IV_{i}$, $t_i$, and $PK_{i+1}$ to be hashed into $h_i$ (\textbf{integrity and freshness}).
    \[
    C_i = AES256Enc(IV_{i}, k_{i}, F_i)
    \]
    \[ 
    h_i = HMAC(k_{i}, C_i||IV_{i}||t_i||PK_{i+1})
    \]
    \item A signature $\sigma_i$ is then computed (\textbf{authenticity and integrity}). For the first signature $\sigma_1$, we use the private key $SK_{c}$ of the camera, while for the other frames we use the one-time keys.
    \[
    \sigma_i = \begin{cases*}
      Sign(SK_{c}, h_1) & if $i = 1$ \\
      Sign(SK_{i}, h_i)      & otherwise
    \end{cases*}
    \]
    \item The encrypted and authenticated frames $\left< C_i,IV_{i},t_i,\sigma_i, PK_{i+1} \right>$ are uploaded to the cloud.
\end{enumerate}

Each user who has access to the correct decryption keys can download these encrypted and authenticated frames $\left< C_i,IV_{i},t_i,\sigma_i, PK_{i+1} \right>$. The user can verify the integrity, authenticity, and freshness of the data, then decrypt the frames, and rebuild the video.

\begin{enumerate}
    \item Each cipher $C_i$, encrypted using the initialization vector $IV_{i}$, with timestamp $t_i$, one-time public key $PK_{i+1}$, and signature $\sigma_i$ is downloaded on demand. 
    
    \item The corresponding symmetric key material $k_{t_i}$ is extracted from the key rotation scheme $\mathcal{K}$. The hash $h_i$ of each cipher $C_i$ is computed.
    \[
    k_{i} = Extract(\mathcal{K}, i)
    \]
    \[
    h_i = HMAC(k_{i}, C_i||IV_{i}||t_i||PK_{i+1})
    \]
    \item The signature $\sigma_i$ is verified with the public key $PK_{c}$ of the camera if this is the first frame or with the one-time public key $PK_{i}$ (\textbf{authenticity, integrity, and freshness}).
    \[
    1 \overset{?}{=} \begin{cases*}
      Verify(PK_{c}, \sigma_1, h_1) & if $i = 1$ \\
      Verify(PK_{i}, \sigma_i, h_i)      & otherwise
    \end{cases*}
    \]
    
    \item If the signature is correct, each cipher $C_i$ is then symmetrically decrypted into the corresponding frame $F_i$ to rebuild the video (\textbf{confidentiality}).
    \[
    F_i = AES256Dec(IV_{i}, k_{i}, C_i)
    \]

\end{enumerate}

\section{Protocol Details}
\label{Appendix:protocol}

\subsection{Delegation}
\label{Appendix:delegation}
\ifrevision \hypertargetblue{mr3a}{\mr{3a}}\fi

\begin{table}[h]
\centering
\begin{tabular}[t]{m{0mm}cccc}
\hline
 & Owner & \faEye & \faBluetoothB & Delegatee\\ \hline
 
\multirow{3}{*}{1} & \multirow{3}{*}{\includegraphics[height=3\baselineskip]{Figures/qr_owner.png}} & \multirow{3}{*}{$\overset{ h_{PK_o}}{\longrightarrow}$} & &\\ 
 &&&&\\
 &&&&\\

\multirow{4}{*}{2} & & & \multirow{4}{*}{$\overset{ PK_o}{\longrightarrow}$} & \multirow{2}{*}{$h'_o = hash(PK_o)$}  \\ 
 & & & & \multirow{2}{*}{$h'_o\overset{?}{=} h_{PK_o}$} \\
&&&&\\
&&&&\\

 \multirow{2}{*}{3} & \multirow{2}{*}{$h'_d= hash(PK_d)$} & &  \multirow{2}{*}{$\overset{ PK_d}{\longleftarrow}$}  &  \multirow{2}{*}{$gen(SK_{d}, PK_{d})$}\\
 &&&&\\
 
 \multirow{3}{*}{4} & \multirow{3}{*}{$h'_d \overset{?}{=} h_{PK_d} $} & \multirow{3}{*}{$\overset{ h_{PK_d}}{\longleftarrow}$} & &\multirow{3}{*}{\includegraphics[height=3\baselineskip]{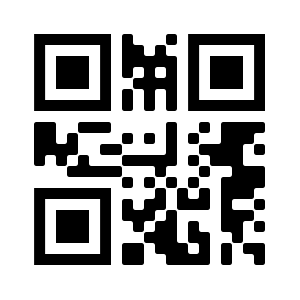}} \\ 
 &&&&\\
 &&&&\\

  \multirow{2}{*}{5} & \multirow{2}{*}{$DH(SK_{o},PK_{d})$} & &  \multirow{2}{*}{$\overset{verif}{\longleftrightarrow}$} & \multirow{2}{*}{$DH(SK_{d}, PK_{o})$}\\ 
 &&&&\\

 \multirow{2}{*}{6} & \multirow{2}{*}{$keys~\{k_{i}\}$} &  & \multirow{2}{*}{$\underset{(RSA)}{\overset{\{k_{i}\}}{\longrightarrow}}$}
 & \multirow{2}{*}{$\mathcal{K} = init(\{k_{i}\})$}\\ 
 & & & & \\
 
\hline
\end{tabular}
\caption{Protocol followed by the smartphone application of the owner and the delegatee during delegation, \faEye{} and \faBluetoothB{} respectively correspond to what is obtained through the visual and Bluetooth channels.}
\label{tab:delegation}
\end{table}

\autoref{tab:delegation} shows the details of the operations performed during the delegation protocol between the owner and a delegatee. The process is very similar to the one done during initialization with the camera:

\begin{enumerate}
    \item The pair of asymmetric keys $(SK_{o}, PK_{o})$ is present on the smartphone of the owner, when the delegation process starts the smartphone application of the owner displays a QR code in which the hash of the owner's public key $h_{PK_o}$ is embedded. The delegatee's smartphone scans this QR code and stores its content.
    
    \item The delegatee and the owner's smartphone connect through Bluetooth and the owner sends its public key $PK_{o}$ to the delegatee, who computes the hash of the owner's public key $h'_o$ and checks that it matches the hash retrieved from the QR code. 
    
    \item If they match, the delegatee's smartphone generates its own asymmetric pair of keys $(SK_{d}, PK_{d})$ and sends its public key $PK_{d}$ through Bluetooth, the owner computes the hash $h'_d$ of the delegatee's public key.
    
    \item The delegatee points their smartphone's screen at the owner. On the screen is displayed the QR code with the hash of the delegatee's public key. The owner retrieves the content from the QR code and checks that it matches $h'_d$.
    
    \item If the key hashes match, both devices now verify that the other device knows the secret key corresponding to the public key that they advertised earlier. This is can be done for instance by applying the Diffie-Hellman key exchange to compute their shared secret and then by exchanging a series of encrypted messages where both parties prove their knowledge.
    
    \item Then, the owner extracts the keys $\{k_i\}$ and share them in an encrypted and authenticated way using RSA with the delegatee to give them access to the corresponding videos. Note that this last step does not necessary need to be done through Bluetooth and could be done over the Internet too without undermining the security of the delegation protocol.

\end{enumerate}

\subsection{Deletion and Factory Reset}
\label{Appendix:deletion}
\ifrevision \hypertargetblue{mr3b}{\mr{3b}}\fi

\begin{table}[h]
\centering
\begin{tabular}[t]{m{0mm}ccc}
\hline
 & Owner & \faBluetoothB{} or \faGlobe & Camera\\ \hline
 
\multirow{4}{*}{1} & &\multirow{4}{*}{$\underset{(RSA)}{\overset{\overset{request}{key~material}}{\longrightarrow}}$} & \multirow{2}{*}{Verify request?}\\ 
& &  & \\
 & &  & \multirow{2}{*}{Perform operation}\\
& &  & \\

\hline
\end{tabular}
\caption{Protocol followed by the owner and the camera during deletion. \faBluetoothB{} and \faGlobe{} correspond to what is obtained through Bluetooth or over another channel such as the Internet.}
\label{tab:deletion}
\end{table}

When the owner decides to delete some videos, they delete the corresponding decryption keys in the key tree $\mathcal{K}$ they have access to. They also need to update accordingly the \textit{key material} inside the \textit{escrow material} saved on the camera. Similarly to the last step of the delegation protocol, updating the \textit{key material} on the camera can be done through Bluetooth or remotely over the Internet. \autoref{tab:deletion} shows the details of this part of the protocol: 

\begin{enumerate}
    \item The owner just sends the timestamped and authenticated request as well as the updated \textit{key material} encrypted and authenticated with RSA (recall that the owner and the camera have shared their asymmetric public keys during initialization). At reception, the camera verifies the authenticity of the update order and performs the operation if it is valid. 
\end{enumerate}

Likewise, to factory reset the camera, the owner sends the request to the camera (through Bluetooth or remotely over the Internet), timestamped and authenticated with the secret key $SK_o$ of the owner to verify the legitimacy of the request. Then, both the camera and owner's smartphone delete the key tree $\mathcal{K}$ they have access to, returning both devices to an uninitialized state. 

\subsection{Access Recovery}
\label{Appendix:recovery}
\ifrevision \hypertargetblue{mr3c}{\mr{3c}}\fi

\begin{table}[h]
\centering
\begin{tabular}[t]{m{0mm}ccc}
\hline
 & New smartphone & \faBluetoothB & Camera\\ \hline

\multirow{2}{*}{1} & &\multirow{2}{*}{$\overset{ request}{\longrightarrow}$} & \\ 
 & &  & \\

\multirow{2}{*}{2} & \multirow{2}{*}{Passphrase known?} &  \multirow{2}{*}{$\underset{material}{\overset{escrow}{\longleftarrow}}$}  & \\
&&&\\

\multirow{2}{*}{3} & $(SK_o,PK_o)$, $PK_c$, &   & \\
& and $\mathcal{K}$ retrieved&&\\

\hline
\end{tabular}
\caption{Protocol followed by the camera and the smartphone application of the user trying to recover access. \faBluetoothB{} correspond to what is obtained through the Bluetooth channel.}
\label{tab:recovery}
\end{table}

\autoref{tab:recovery} shows the steps executed when someone tries to recover access to the system: 

\begin{enumerate}
    \item The owner uses their new smartphone to open a Bluetooth connection to the camera and request the escrow material.
    
    \item The camera sends back the escrow material. Recall that this escrow material is encrypted with the recovery passphrase that was displayed to the owner during initialization. 
    
    \item If the owner knows the recovery passphrase, they are able to recover access to the asymmetric key pair $(SK_o,PK_o)$ of the owner, to the public key $PK_c$ of the camera, and to the \textit{key material} necessary to build the key tree $\mathcal{K}$.

\end{enumerate}

\section{Functional User Evaluation Protocol}
\label{Appendix:user_protocol}

All institutional requirements were met for this functional user evaluation of \cactus{}. We obtained approval from the Institutional Review Board (IRB) of our university and a consent form was signed by the participants at the beginning of their session. We also tested the protocol with coworkers and collaborators beforehand to identify possible limitations. 

All the material was provided to the participants that were guided by a researcher through the different tasks to perform. We introduced each task with a real-life scenario to help the participants behave as if they were using the system in their real life. To collect feedback, we observed a talk aloud process asking the participants to express aloud what they are doing or looking for while performing the task, allowing us to better identify potential issues in the system. Between each task, we also asked specific questions about the process that had just been completed.

\subsection{Before Each Session}

Before each session, we verified that all the material needed for the session was provided, was working as expected, and was in its initial state:

\begin{enumerate}
    \item \textbf{Camera Device:} 1 \textit{Raspberry Pi 4 Model B  2GB}, 1 camera sensor, 1 case, 1 power supply, 1 micro SD card with OS image flashed on it and our software installed.
    \item \textbf{Android Smartphone Devices:} 3 \textit{Nokia 4.2} with our application pre-installed, wifi connection configured, location enabled (to enable discovery of nearby Bluetooth devices), Bluetooth disabled with no prior device paired, and with no saved media on the smartphone.
    \item \textbf{Other:} Piece of paper and pen provided (to write down the recovery passphrase).
\end{enumerate}

\subsection{During Each Session}

As participants are being recorded using the system and are recording their own video, we first asked participants for explicit consent to be recorded. We then proceeded by asking some preliminary questions about their background and their experience with smart camera systems to get the discussion started.

Next, we explained that the objective of the session was to evaluate the functional usability of the smart camera system that we designed. Details about privacy violations in current available systems were briefly described to help the participants understand the motivation of the project. Then, we presented how the system we implemented was giving back full control to the users and enforcing their privacy. These explanations were very high level, since our goal was to make sure that the participants understood that the video frames were encrypted before leaving the camera device and being uploaded to cloud storage, that they were then decrypted on the smartphone side, and that the encryption and decryption keys were only known by the devices that the user would allow to share access with.

After describing how the rest of the session was going to take place, each task was introduced by a scenario and feedback obtained through the talk aloud process as well as follow up questions. We verified that the participants were either familiar with Android or we showed them how to navigate between applications on Android.

Finally, we asked each participant to perform the different tasks in the following order.

\vspace{-1\baselineskip}
\subsubsection{Initialization}

\textbf{Scenario:} You want to secure your home, so you just bought this new smart camera system online. You received the package with the camera and just downloaded the application on your smartphone. Go ahead with the rest of the configuration. \\
\textbf{Questions:}

\begin{itemize}
    \item What did you like about the initialization? 
    \item What did you dislike about the initialization?
    \item Any further comments?
\end{itemize}

\subsubsection{System Usage}

\textbf{Scenario:} You have your new smart camera all set up, so now you want to be able to see what is happening inside/outside of your home. For that you open the application to view the live streaming and access the different functionalities of the application.  \\
\textbf{Questions:}

\begin{enumerate}
    \item How do you feel about the quality of the video streaming? 
    \item What is your opinion about the following statements (Likert scale: strongly disagree, disagree, neutral, agree, strongly agree)?
    
    \begin{enumerate}
        \item I am happy with the image quality of the streaming.
        \item I am happy with the latency of the streaming.
        \item I am happy with the frame rate of the streaming.
    \end{enumerate}
    
    \item What did you like about the usage of the system/application? 
    \item What did you dislike about the usage of the system/application? 
    \item Any further comments?
\end{enumerate}

\subsubsection{Delegation} 

\textbf{Scenario:} You want to give access to someone else to the live streaming of your camera, as you are going on vacation abroad. They have downloaded the application on their phone, you need to add them as a new delegatee on your application. \\
\textbf{Questions:}

\begin{enumerate}
    \item What did you like about the delegation process? 
    \item What did you dislike about the delegation process?
    \item What is your opinion about the following statement: It is easy to add a new delegatee (Likert scale: strongly disagree, disagree, neutral, agree, strongly agree)?
    \item Would you like to see any change in the delegation process or the options for the access control? 
    \item Who would you typically add as a delegatee and for how long?
    \item What do you think of the granularity of the delegation control?
    \item Any further comments?
\end{enumerate}

\subsubsection{Access Recovery}

\textbf{Scenario:} Unfortunately, on your way back home during the layover, you lost your smartphone, which was the device you used to access your camera system, and when you come back home, you figured out that someone has broken in during your holidays and robbed you. You buy a new smartphone, install back the application on it, and want to recover access to your system to see what happened. Luckily, you wrote down your recovery passphrase.\\
\textbf{Questions:}

\begin{enumerate}
    \item What did you like about the recovery process? 
    \item What did you dislike about the recovery process?
    \item What is your opinion about the following statement: It is important for me to have a recovery process. (Likert scale: strongly disagree, disagree, neutral, agree, strongly agree)?
    \item Any further comments?
\end{enumerate}

\subsubsection{Factory Reset}

\textbf{Scenario:} Finally, you want to reconfigure your system as you lost your smartphone, but before you want to make sure to factory reset the system. \\
\textbf{Questions:}

\begin{enumerate}
    \item What did you like about the factory reset process? 
    \item What did you dislike about the factory reset process?
    \item What is your opinion about the following statement: It is important for me to have a factory reset process. (Likert scale: strongly disagree, disagree, neutral, agree, strongly agree)?
    \item Any further comments?
\end{enumerate}

\subsection{After Each Session}
After each session, we made sure that every device was reset and back into its initial state, as if the session had not occurred.

\section{Storyboard}
\label{Appendix:storyboard}

Following is the storyboard of the \cactus{}'s smartphone application.

\begin{figure}[!ht]
    \centering
    \includegraphics[scale=0.44]{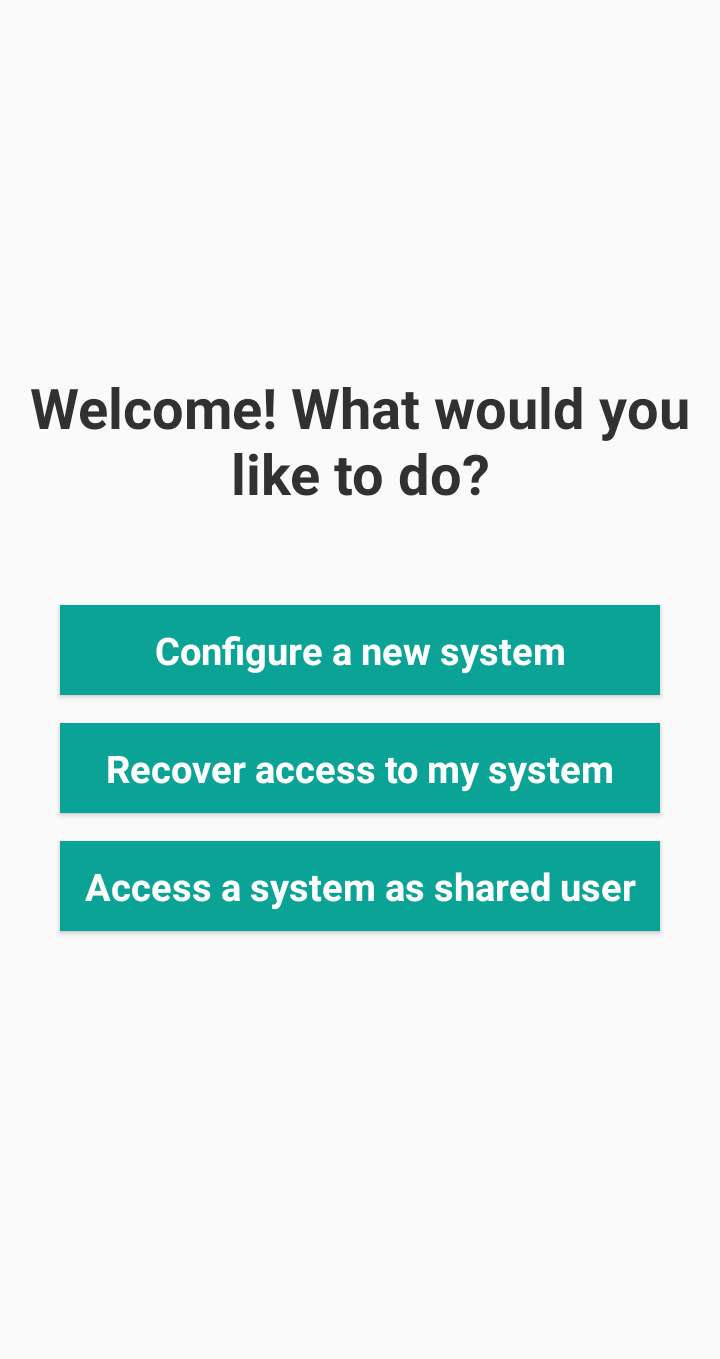}
    \includegraphics[scale=0.44]{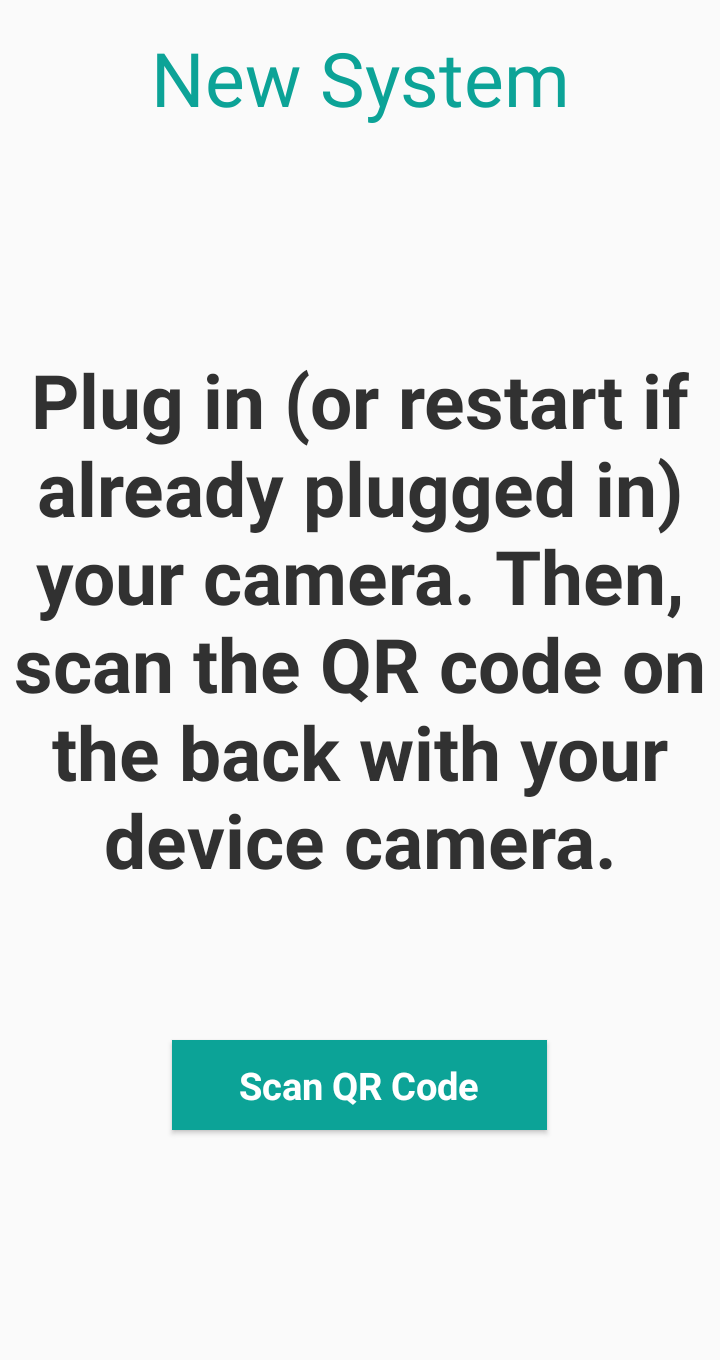}
    \includegraphics[scale=0.44]{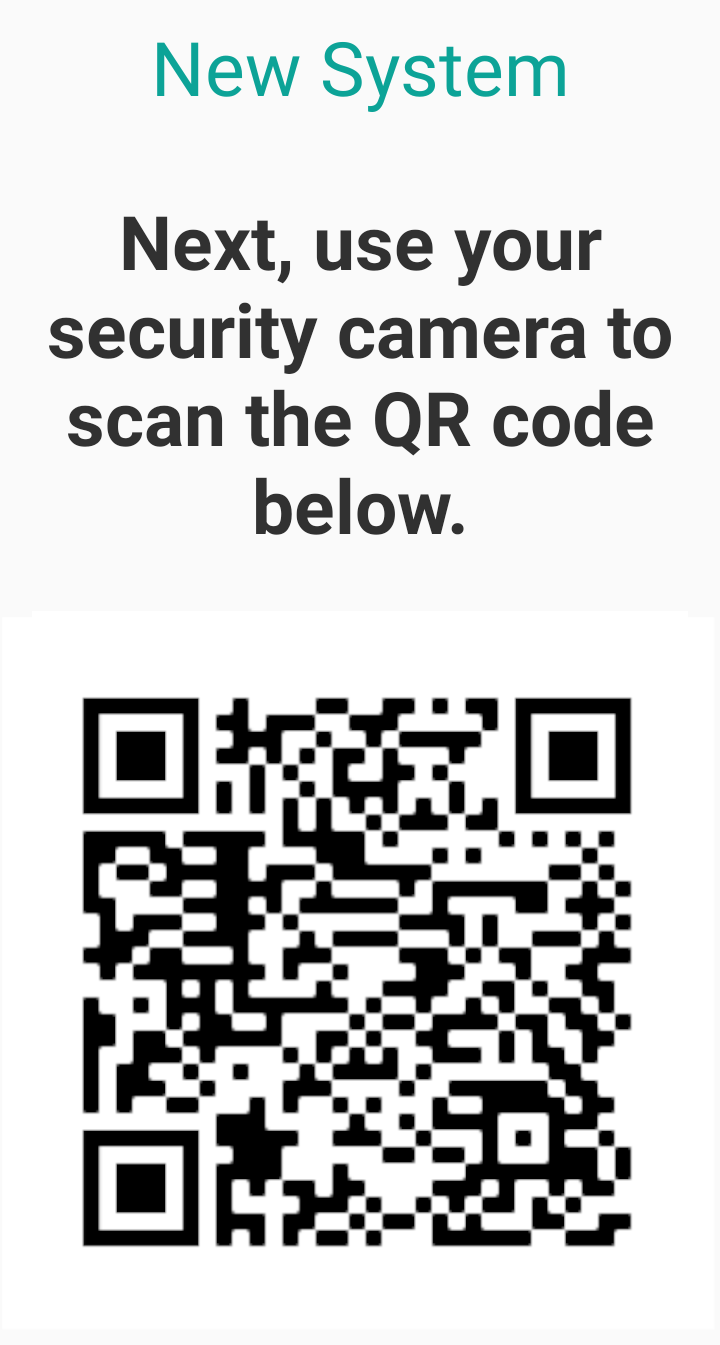}
    \includegraphics[scale=0.44]{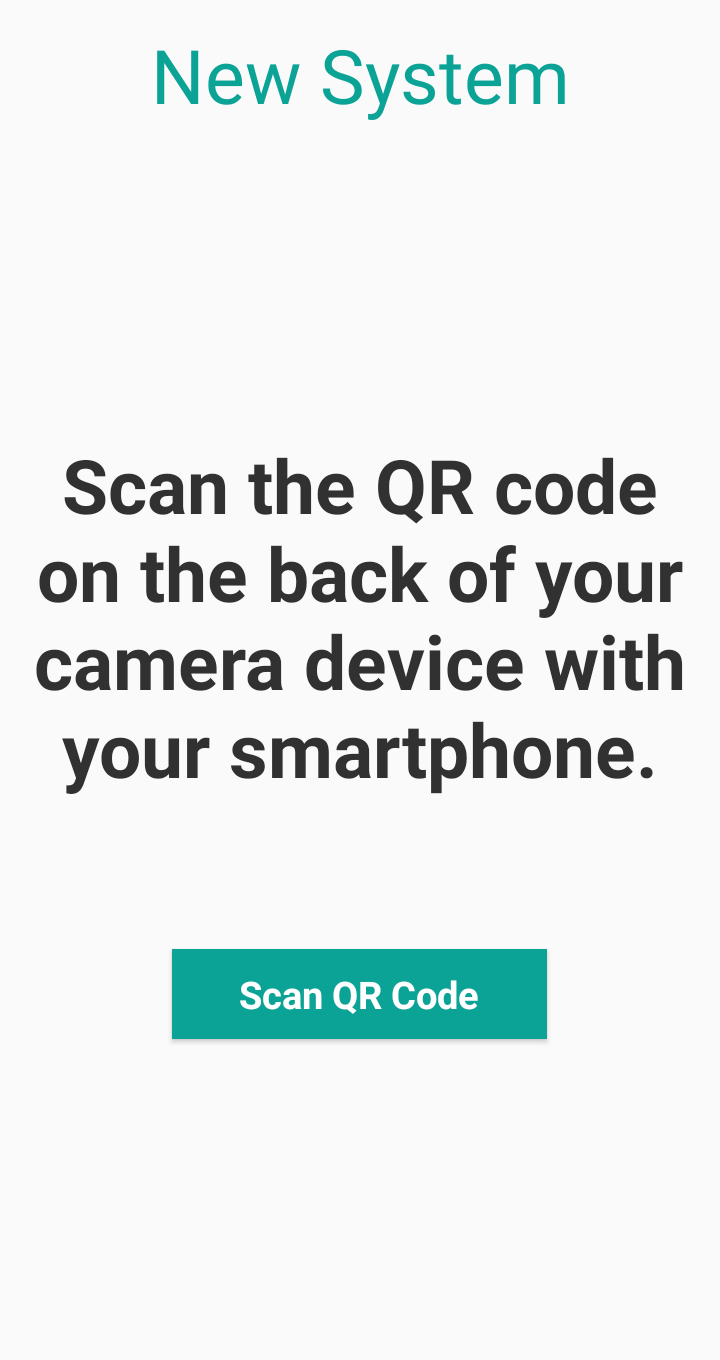}
    \includegraphics[scale=0.44]{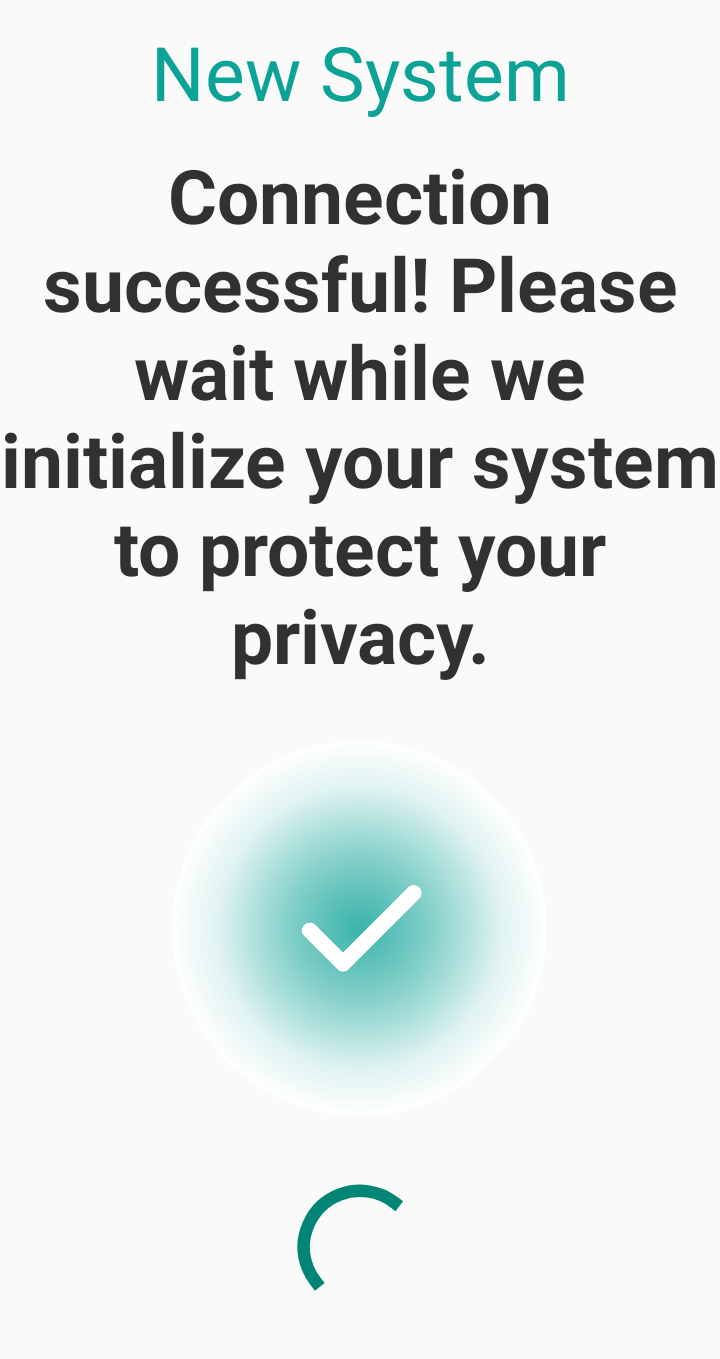}
    \includegraphics[scale=0.44]{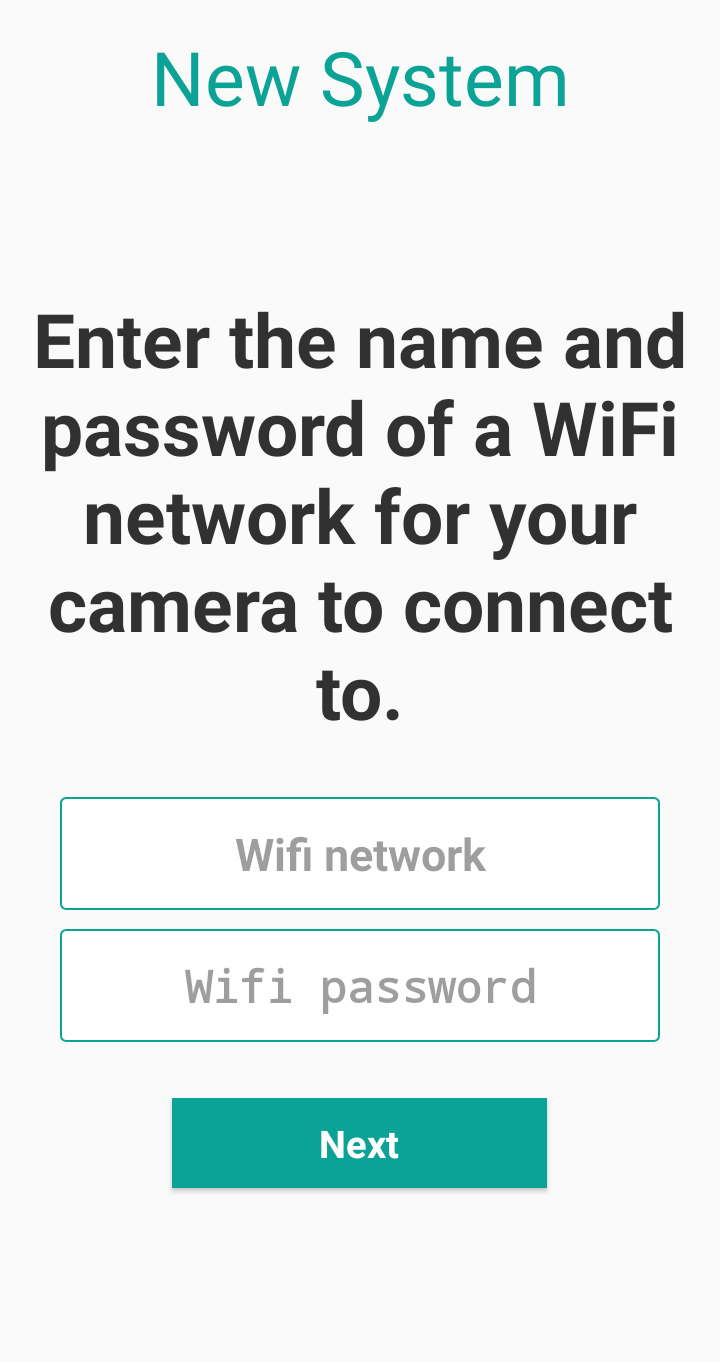}
    \includegraphics[scale=0.44]{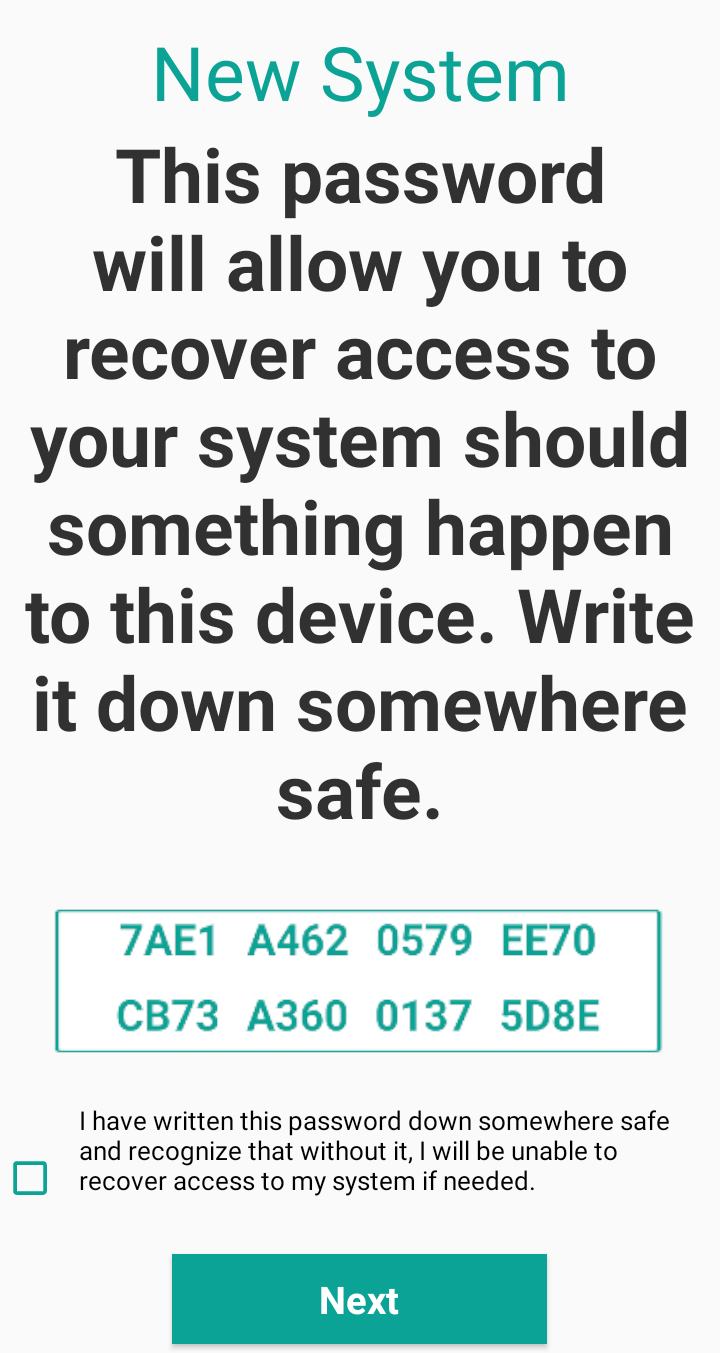}
    \includegraphics[scale=0.44]{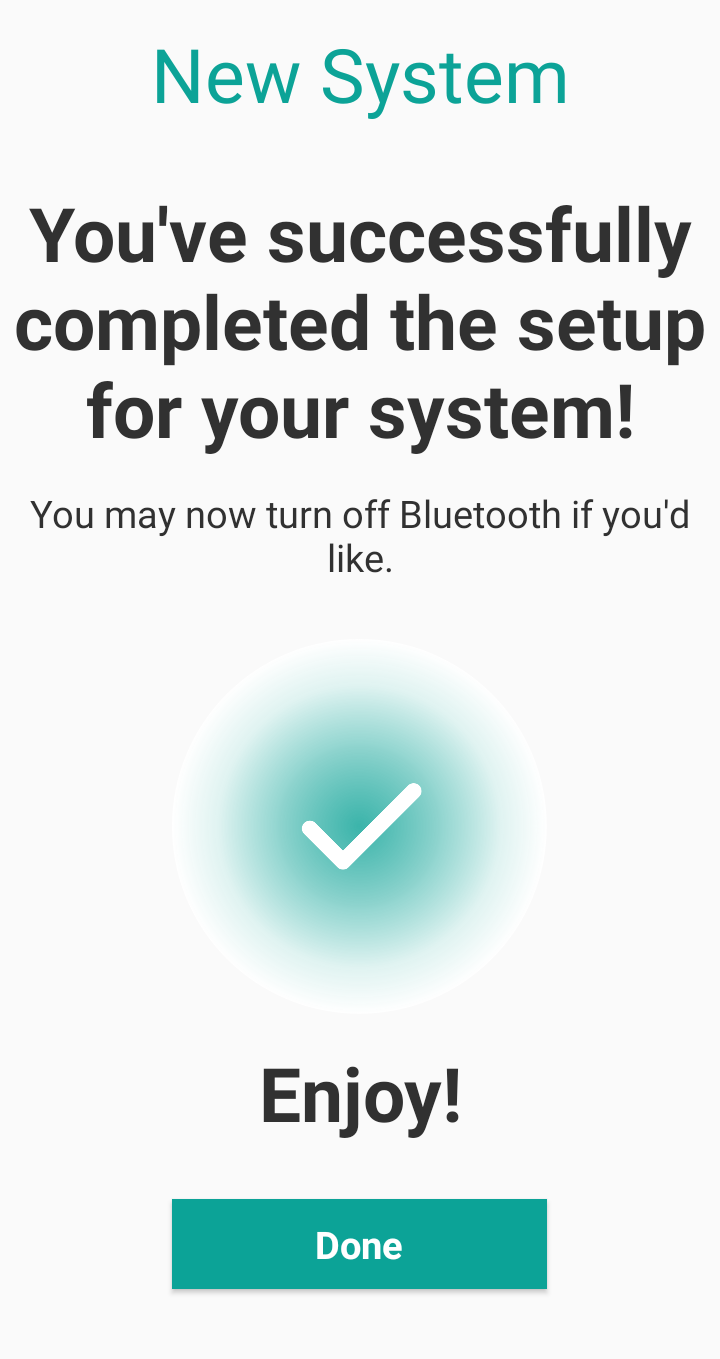}
    \caption{Screenshots of the \cactus{} smartphone application of the owner during initialization of the system.}
\end{figure}

\begin{figure}[!ht]
    \centering
    \includegraphics[scale=0.44]{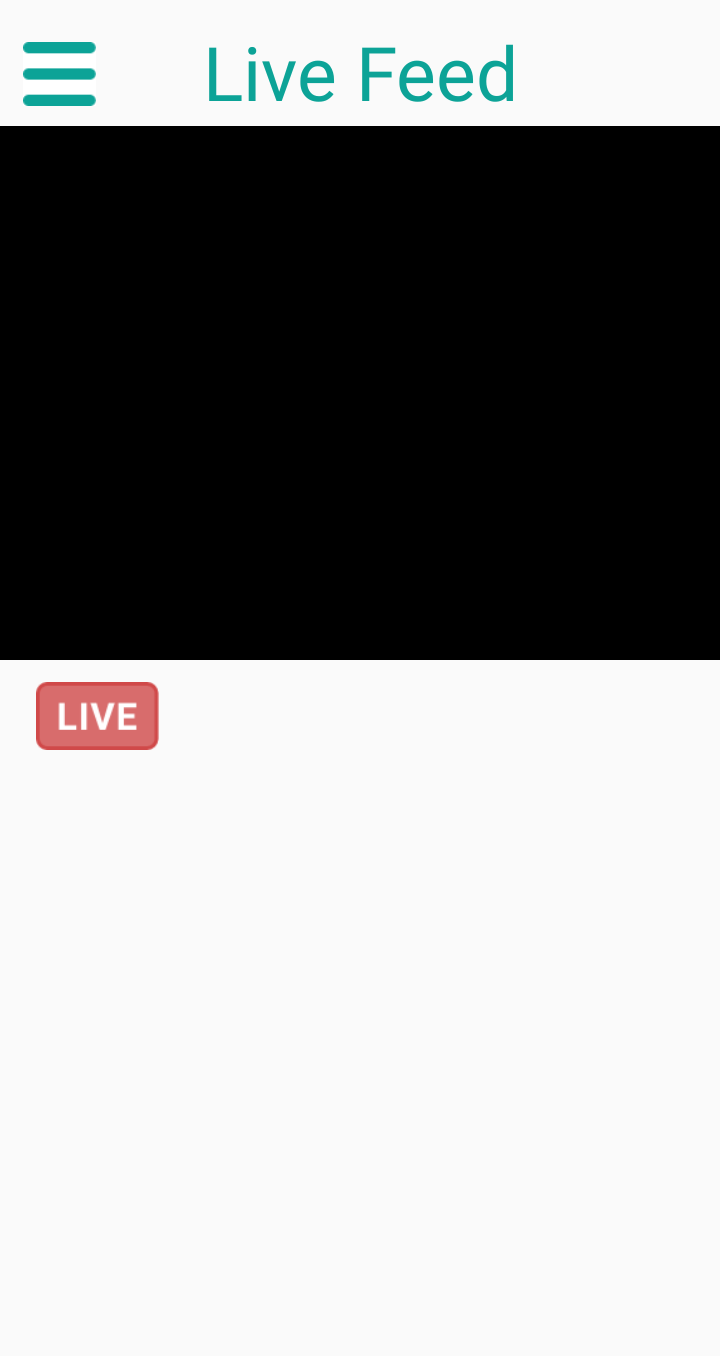}
    \includegraphics[scale=0.44]{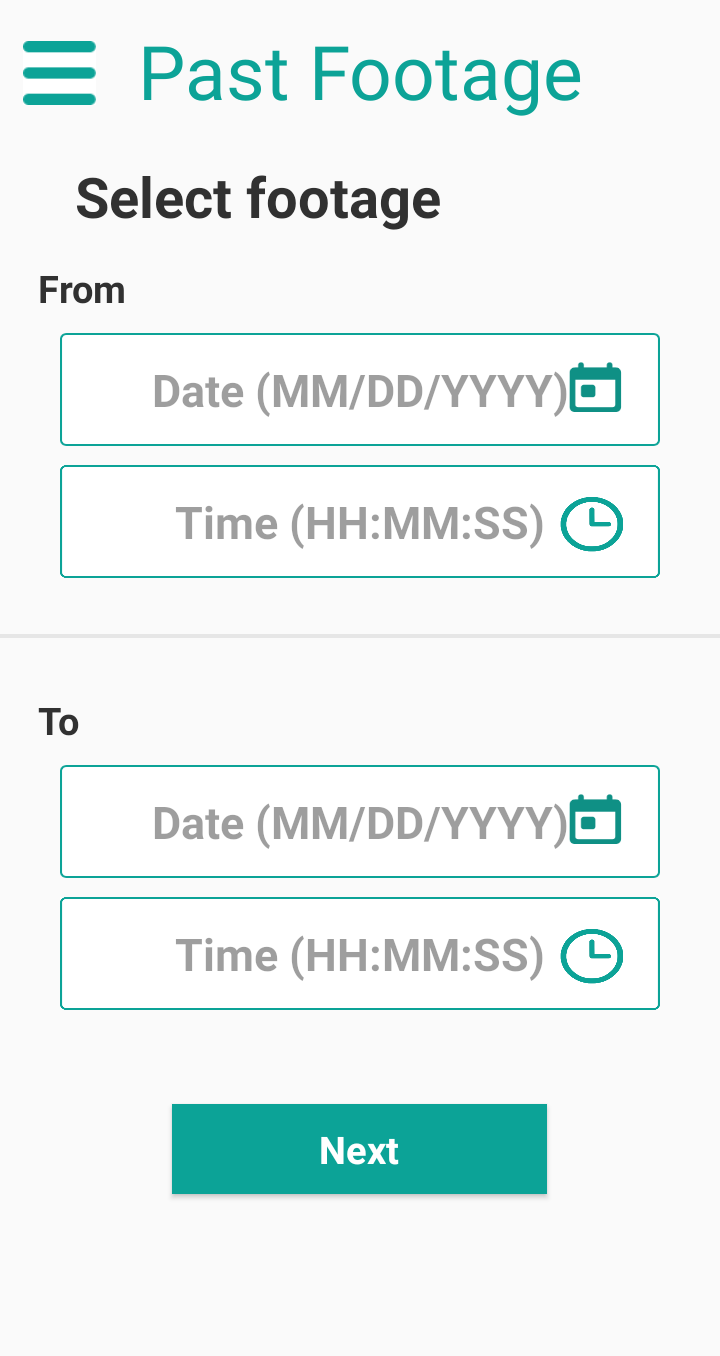}
    \includegraphics[scale=0.44]{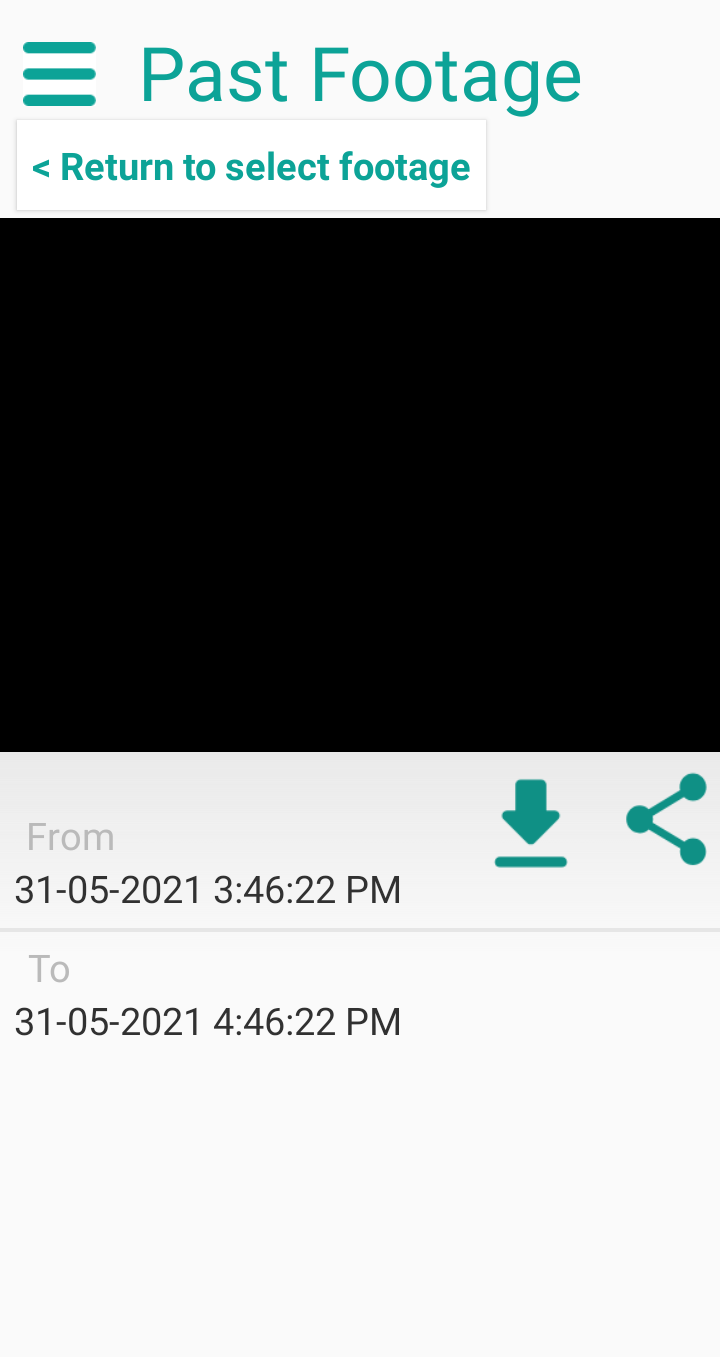}
    \includegraphics[scale=0.44]{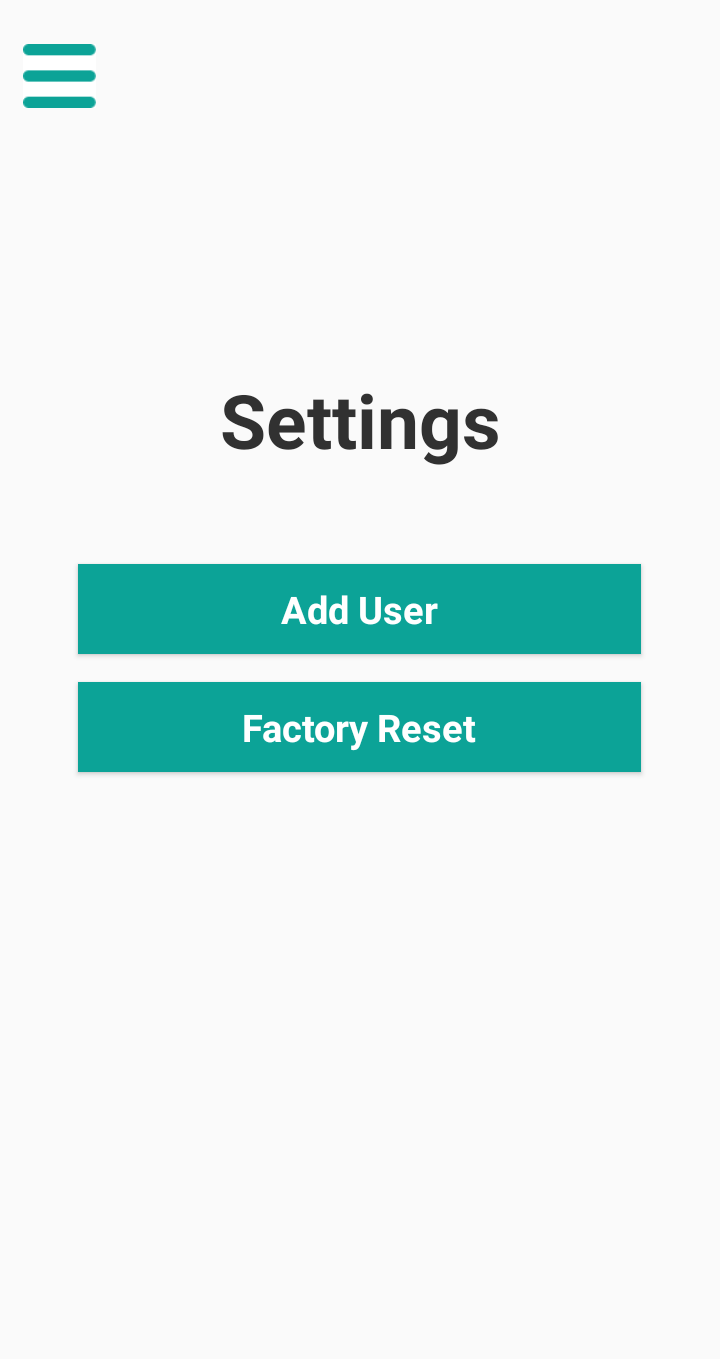}
    \caption{Screenshots of the \cactus{} smartphone application: home page (live feed), access to past footage, and settings.}
\end{figure}

\begin{figure}[!ht]
    \centering
    \includegraphics[scale=0.44]{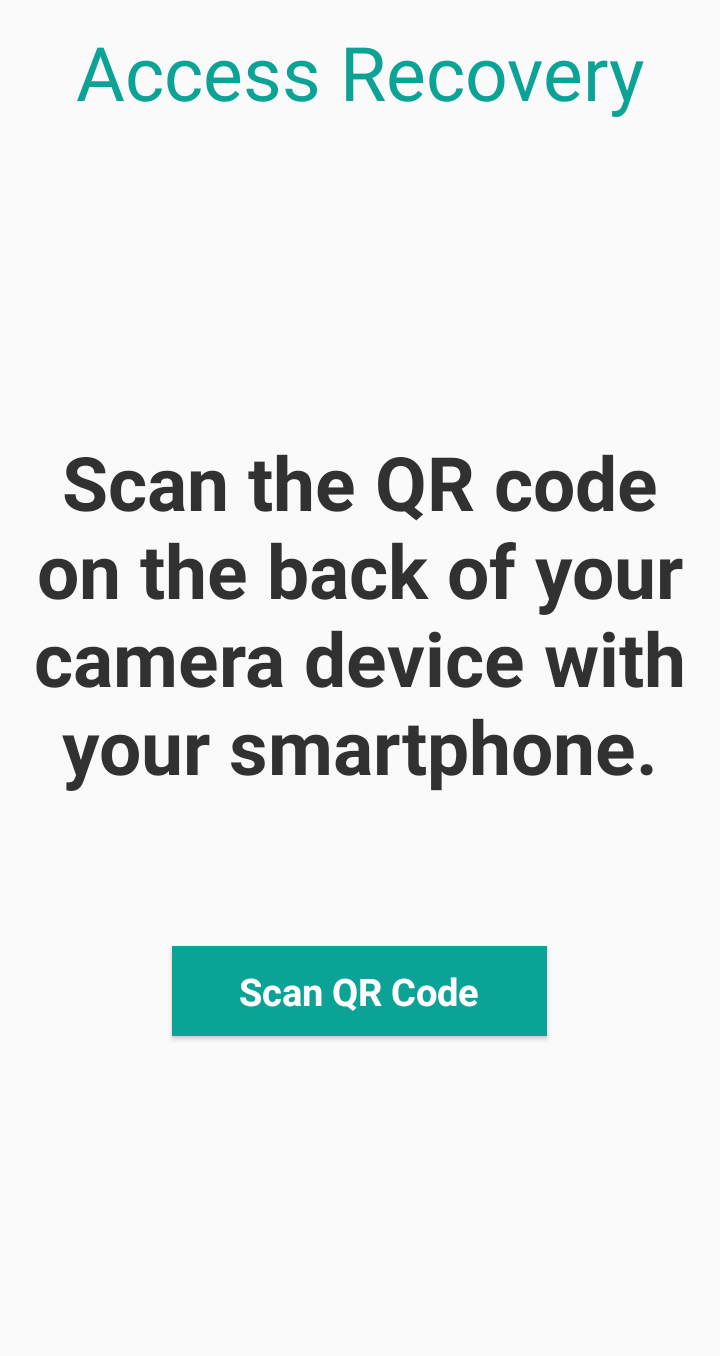}
    \includegraphics[scale=0.44]{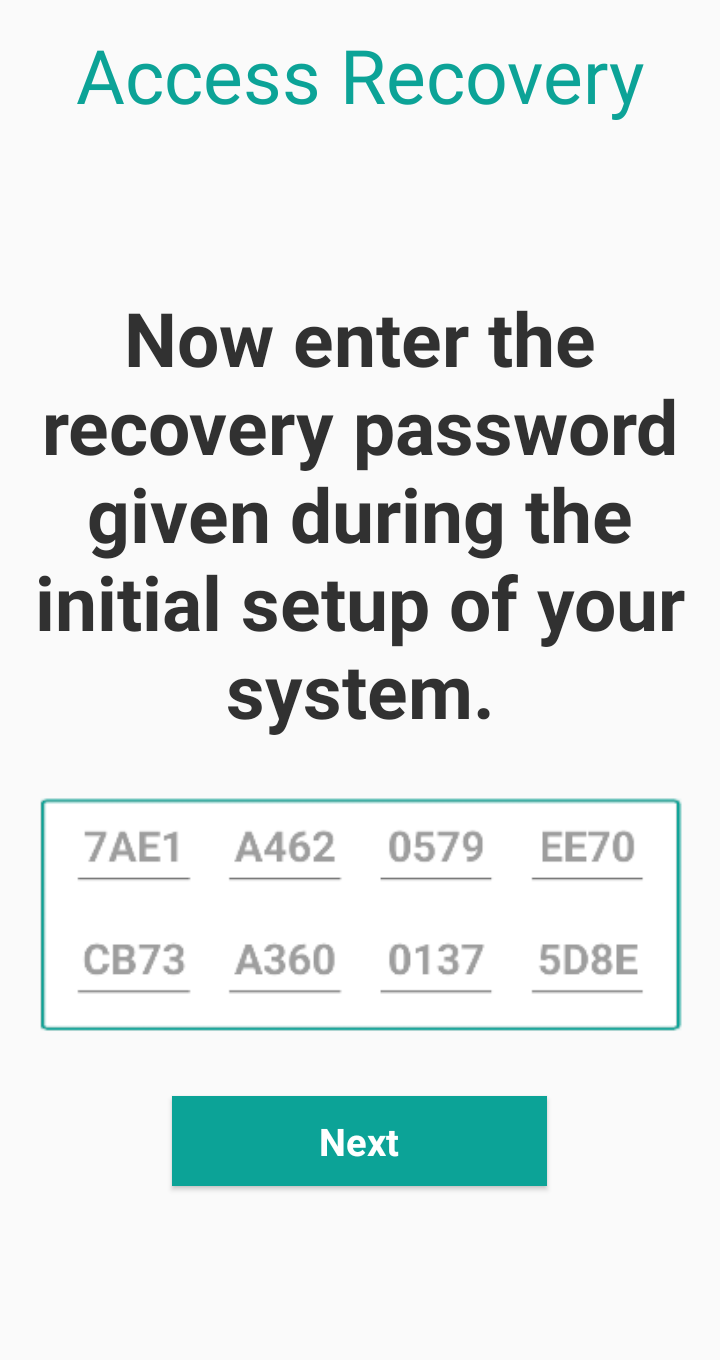}
    \includegraphics[scale=0.44]{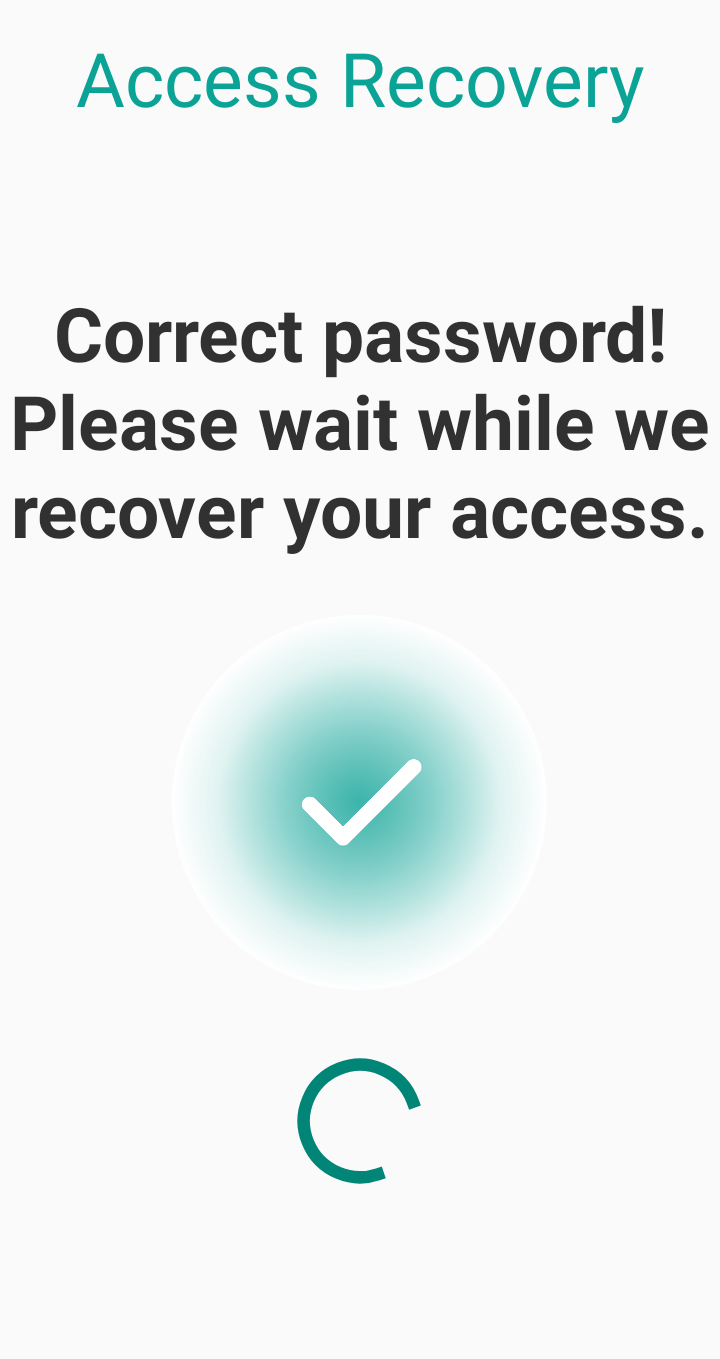}
    \caption{Screenshots of the \cactus{} smartphone application of the owner during access recovery.}
\end{figure}

\begin{figure}[!ht]
    \centering
    \includegraphics[scale=0.44]{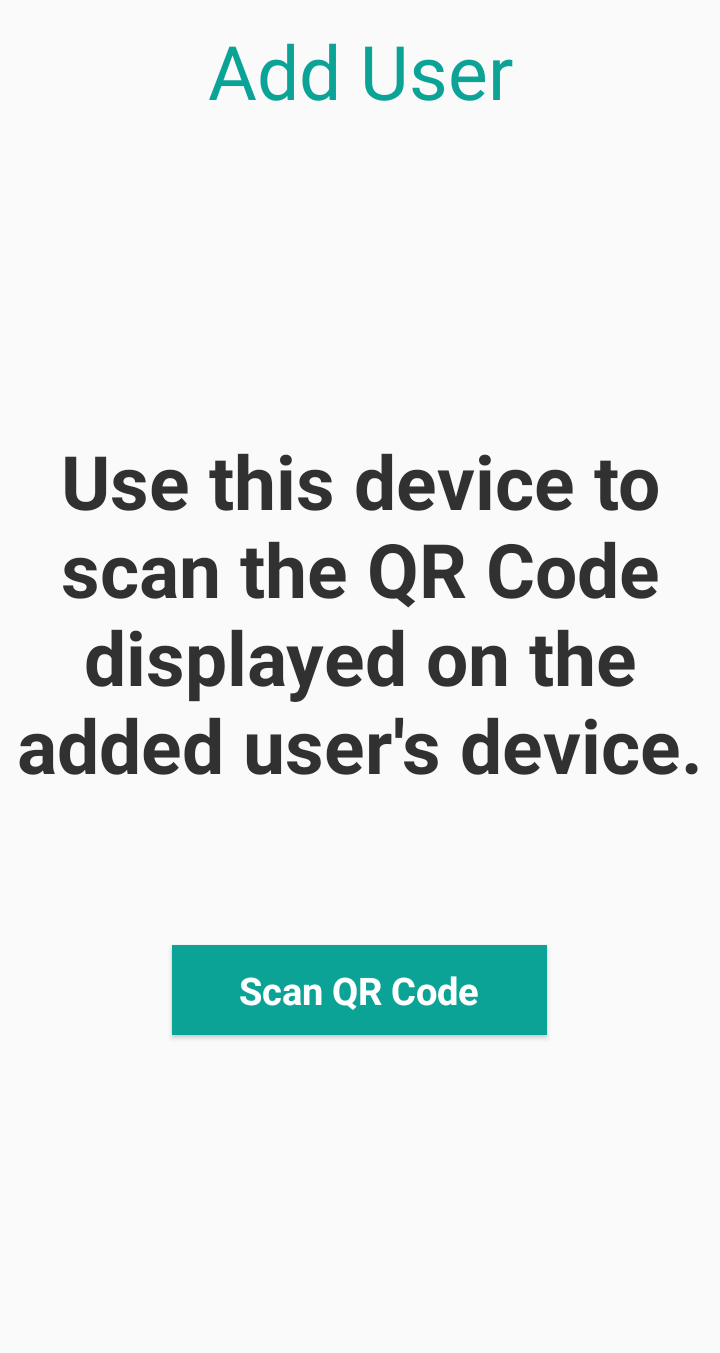}
    \includegraphics[scale=0.44]{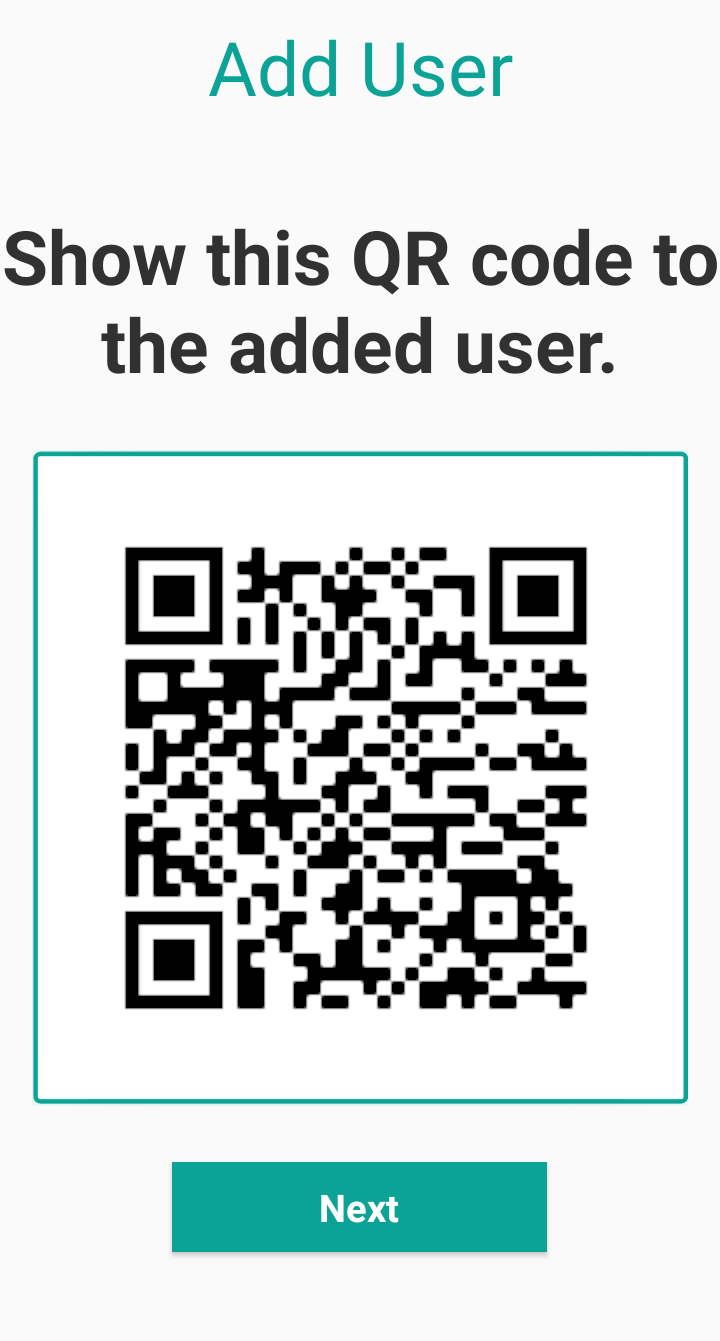}
    \includegraphics[scale=0.44]{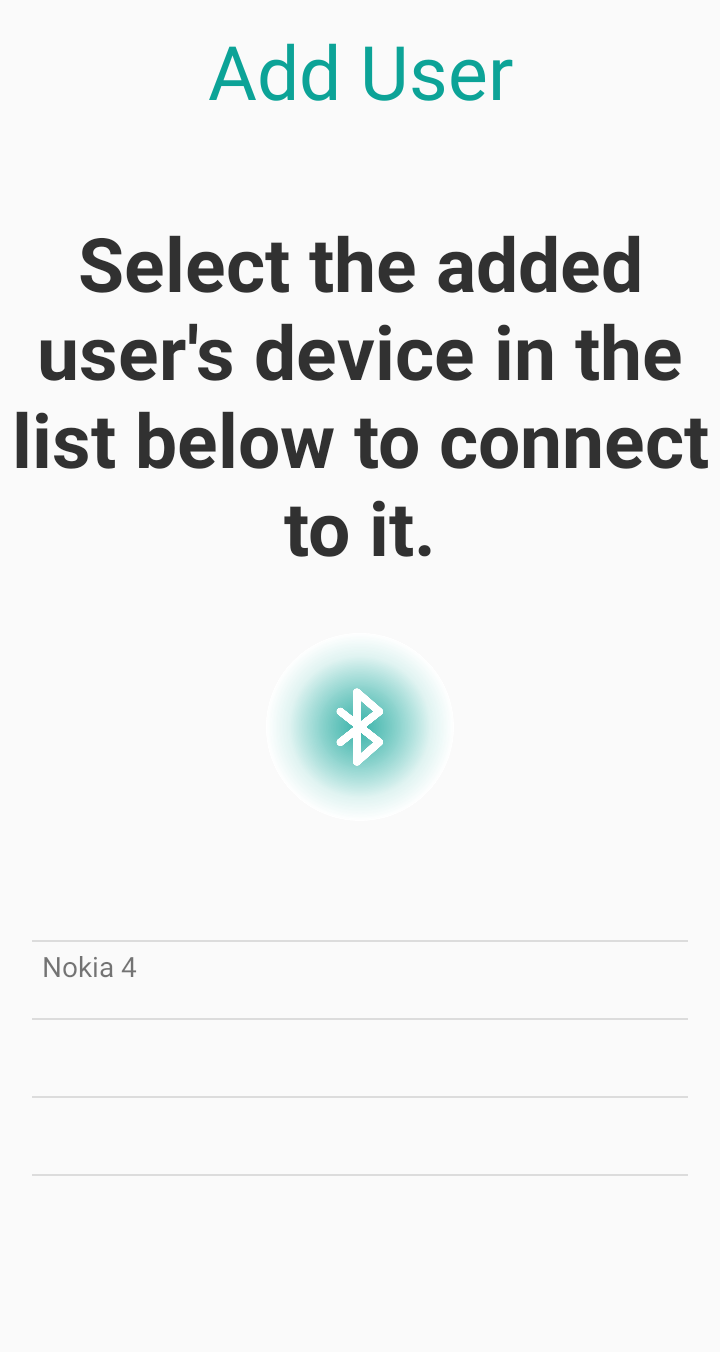}
    \includegraphics[scale=0.44]{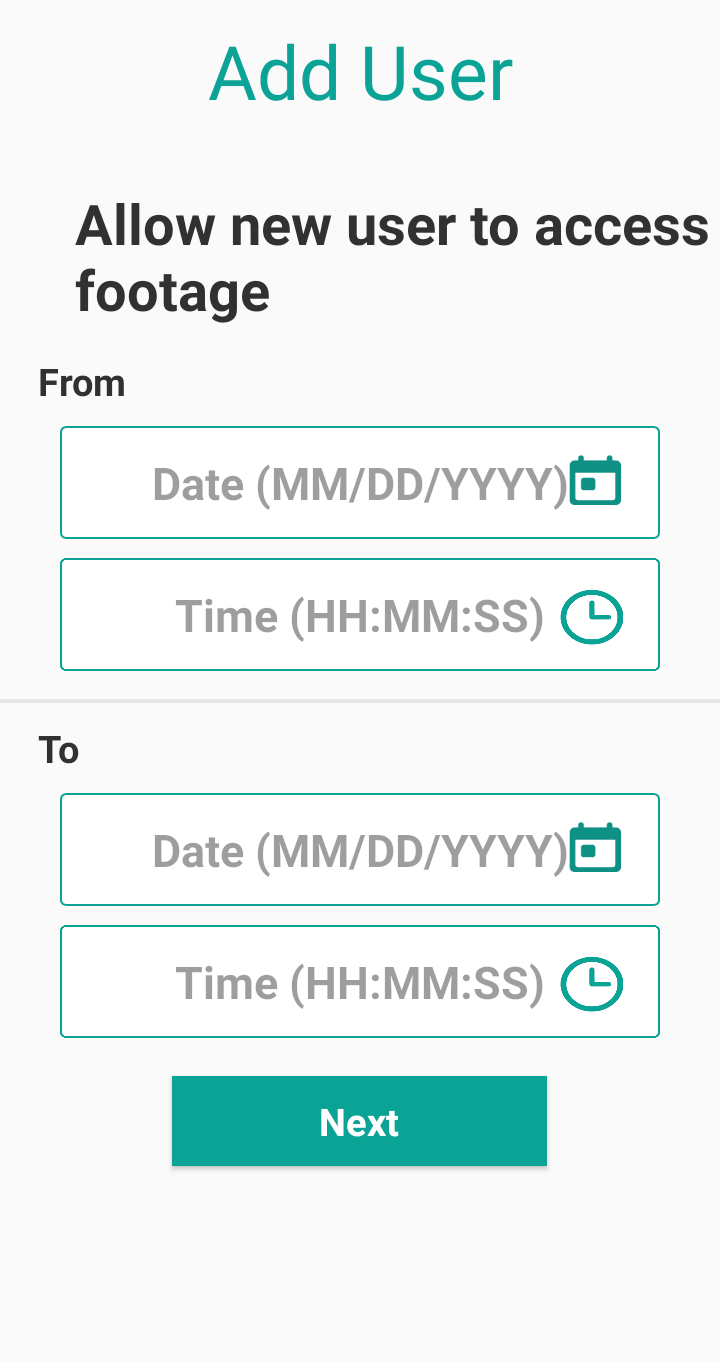}
    \includegraphics[scale=0.44]{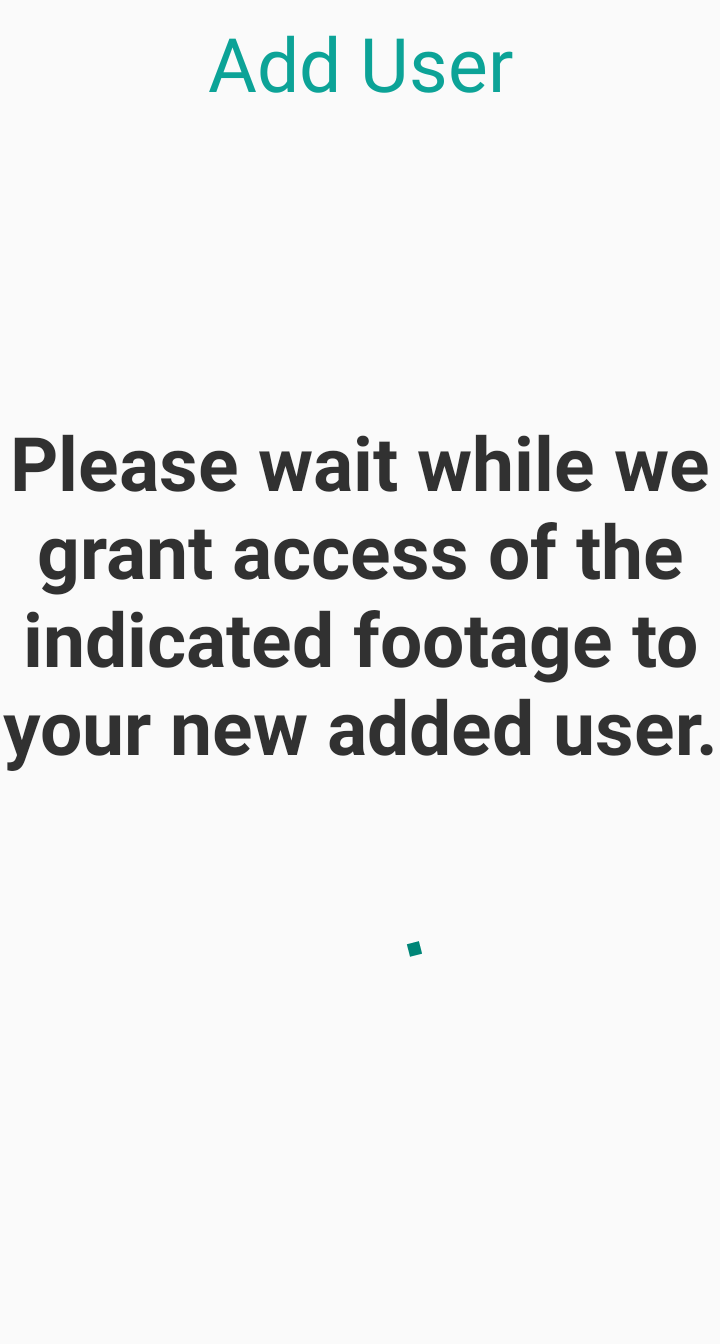}
    \includegraphics[scale=0.44]{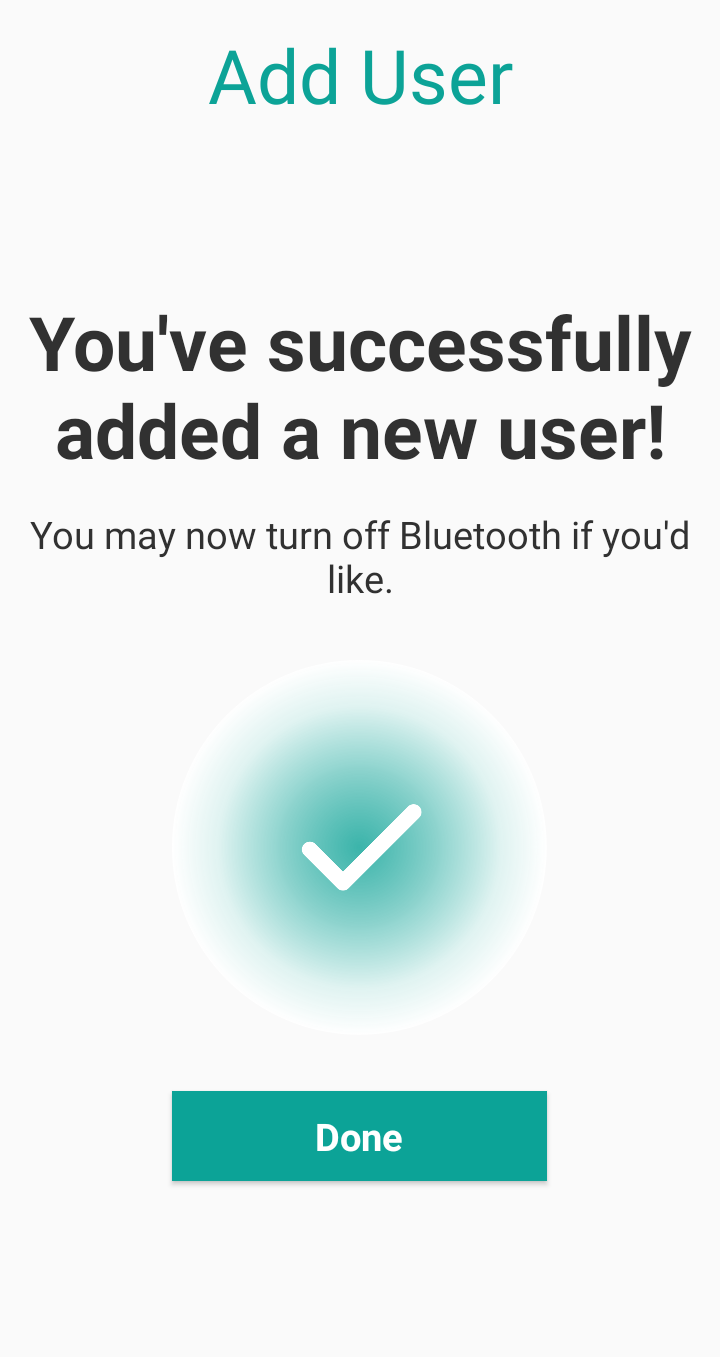}
    \caption{Screenshots of the \cactus{} smartphone application of the owner during delegation.}
\end{figure}

\begin{figure}[!ht]
    \centering
    \includegraphics[scale=0.44]{Figures/screenshots/delegatee/unintialized.png}
    \includegraphics[scale=0.44]{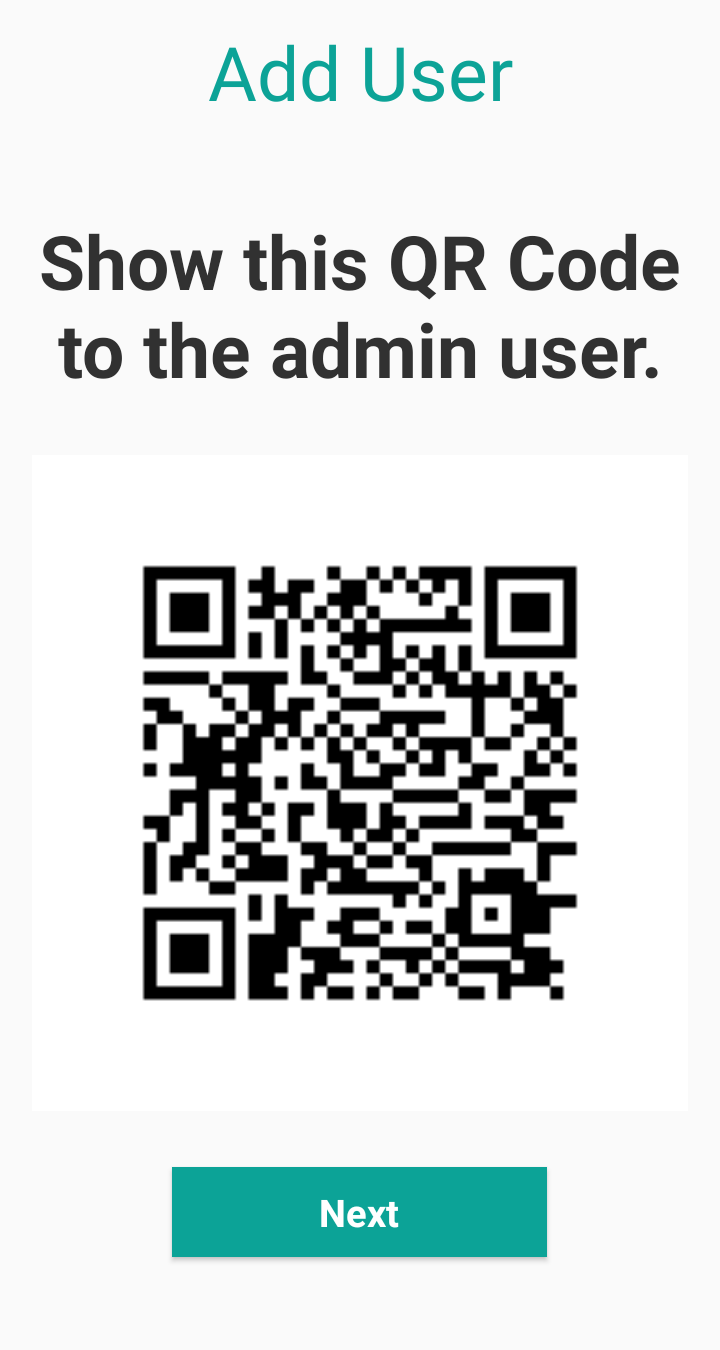}
    \includegraphics[scale=0.44]{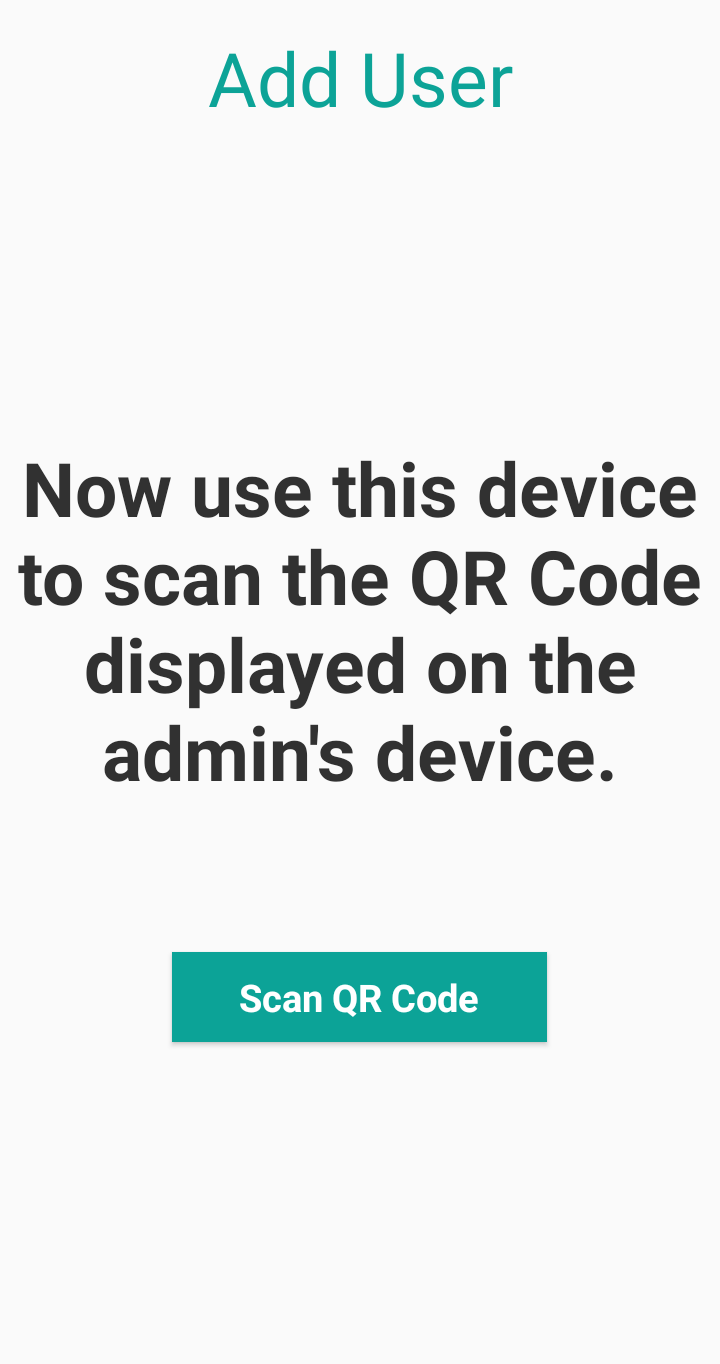}
    \includegraphics[scale=0.44]{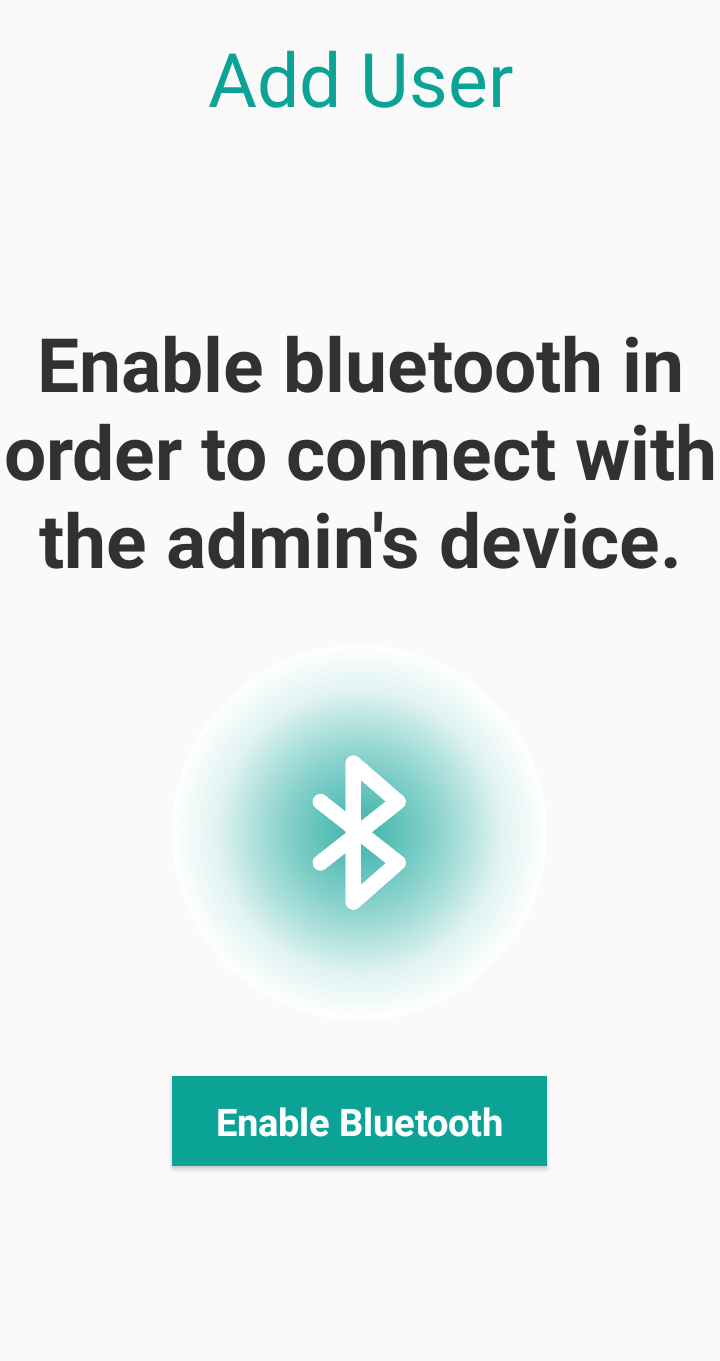}
    \includegraphics[scale=0.44]{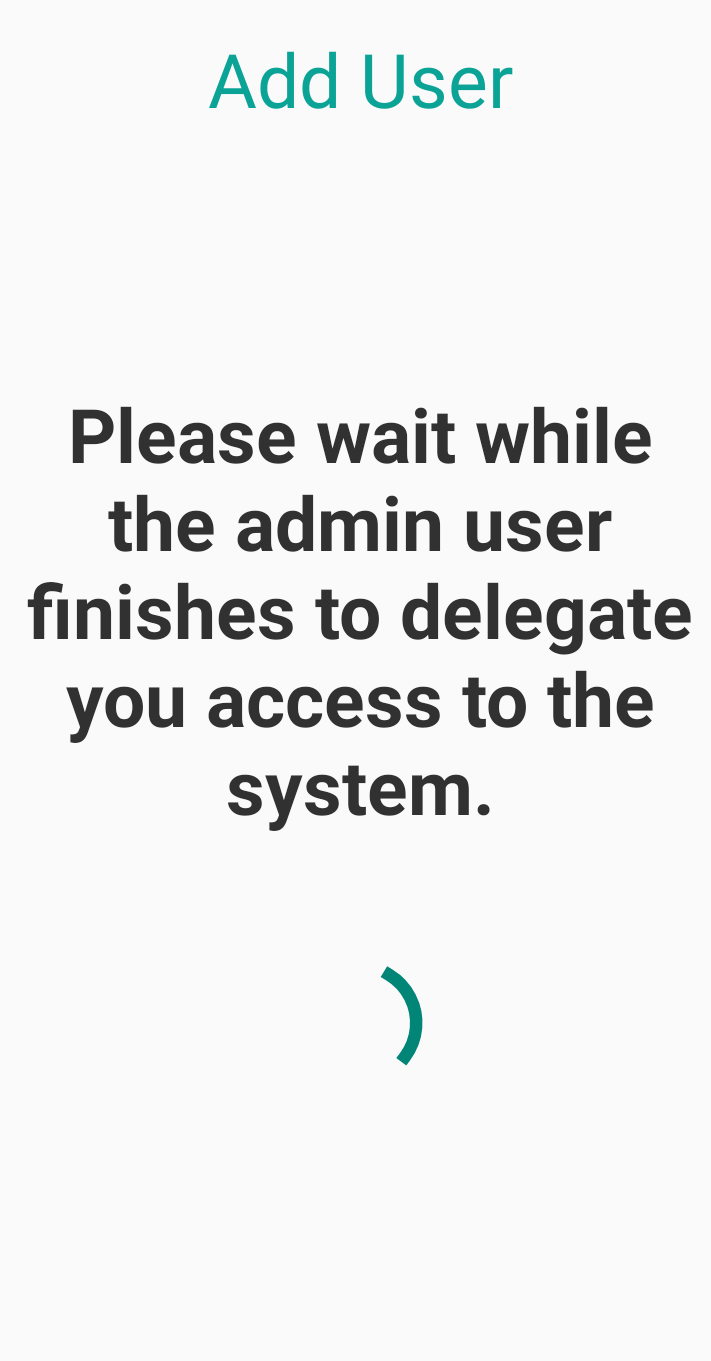}
    \includegraphics[scale=0.44]{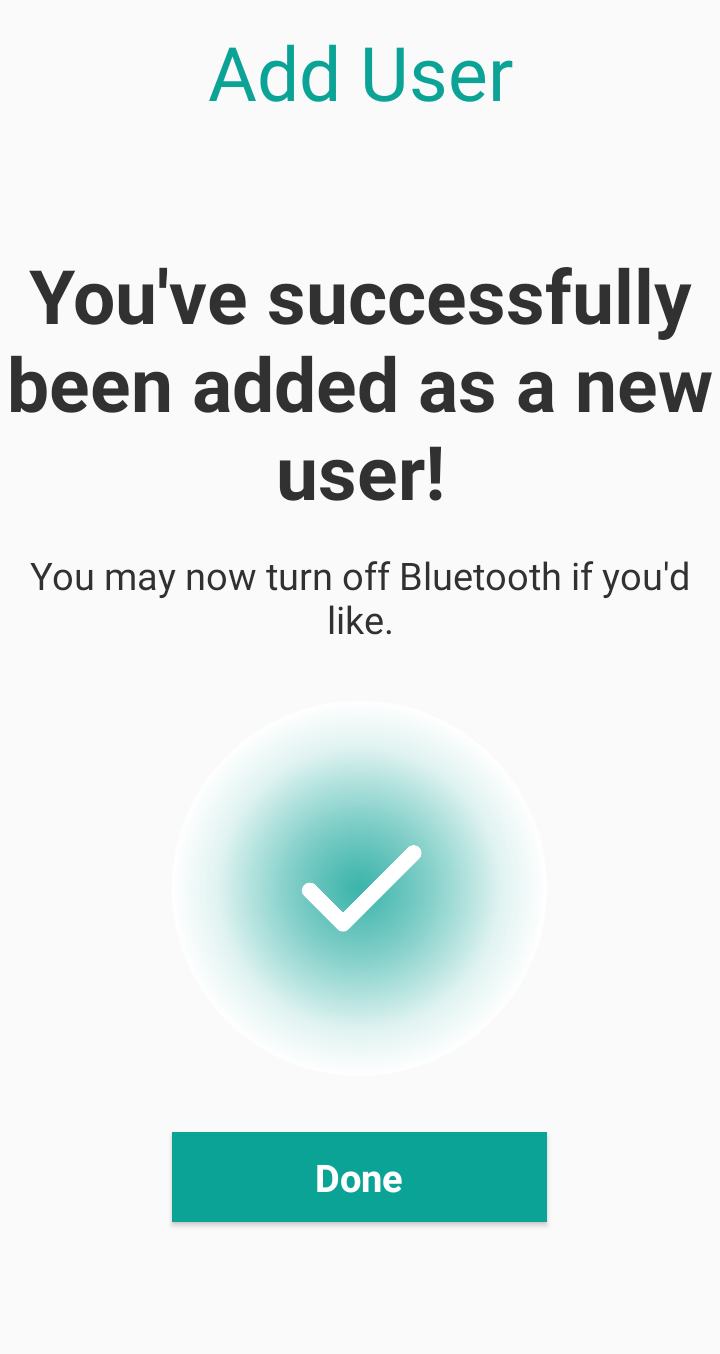}
    \caption{Screenshots of the \cactus{} smartphone application of the delegatee during delegation.}
\end{figure}

%

\end{document}